\documentclass[aps,prb,twocolumn,superscriptaddress,floatfix,showpacs]{revtex4}

\usepackage{graphicx} 
\usepackage{calc} 
\usepackage{amsfonts} 
\usepackage{amsmath} 

\newcommand{\chbem}[1]{}

\newcommand{\glbezug}[1]{(\ref{#1})}

\newcommand{\konjug}{^*}
\newcommand{\uperp}{_{\bot}}
\newcommand{\upara}{_{\|}}

\newcommand{\gnn}{g^{\bot}}
\newcommand{\tildegnn}{\tilde{g}^{\bot}}

\newcommand{\tildevertnn}{{\tilde{\Gamma}}^{\bot}}
\newcommand{\orbvertpart}{\zeta} 

\newcommand{\ww}{g}
\newcommand{\vertex}{\Gamma}

\newcommand{\Cooper}{{\cal C}}
\newcommand{\Landau}{{\cal L}}
\newcommand{\Peierls}{{\cal P}}

\newcommand{\qcooperperp}{q_{{\cal C}\bot}}
\newcommand{\qcooperpara}{q_{{\cal C}\|}}

\newcommand{\qpeierlsperp}{q_{{\cal P}\bot}}

\newcommand{\betrag}[1]{\left| #1 \right|}
\newcommand{\mittelwert}[1]{\langle #1 \rangle}

\newcommand{\signum}{{\rm sgn}}

\newcommand{\zweikleinemathezeilen}[2]{\genfrac{}{}{0pt}{2}{#1}{#2}}

\newlength{\gesamtbildbreite}

\begin{document} 

\title{Superconducting pairing and density-wave instabilities in
  quasi-one-dimensional conductors}        
 
\author{J. C. Nickel}
\affiliation{Regroupement Qu\'ebecois sur les Mat\'eriaux de Pointe,
  D\'epartement de physique, Universit\'e de Sherbrooke, Sherbrooke,
  Qu\'ebec, Canada, J1K-2R1 }
\affiliation{Laboratoire de Physique des Solides, CNRS UMR 8502,
  Universit\'e Paris-Sud, 91405 Orsay, France} 
\author{R.  Duprat}
\affiliation{Regroupement Qu\'ebecois sur les Mat\'eriaux de Pointe,
  D\'epartement de physique, Universit\'e de Sherbrooke, Sherbrooke,
  Qu\'ebec, Canada, J1K-2R1 }
\author{ C. Bourbonnais}
\affiliation{Regroupement Qu\'ebecois sur les Mat\'eriaux de Pointe,
  D\'epartement de physique, Universit\'e de Sherbrooke, Sherbrooke,
  Qu\'ebec, Canada, J1K-2R1 }
\author{N. Dupuis}
\affiliation{Laboratoire de Physique des Solides, CNRS UMR 8502,
  Universit\'e Paris-Sud, 91405 Orsay, France}
\affiliation{Department of Mathematics, Imperial College, 
180 Queen's Gate, London SW7 2AZ, UK}
 
\date{October 25, 2005}

\begin{abstract} 
Using a renormalization group approach, we determine the phase diagram of an
extended quasi-one-dimensional electron gas model that includes interchain
hopping, nesting deviations and both intrachain and interchain repulsive
interactions. $d$-wave superconductivity, which dominates over the
spin-density-wave (SDW) phase at large nesting deviations, becomes unstable to
the benefit of a triplet $f$-wave phase for a weak repulsive interchain
backscattering term $g_1^\perp>0$, despite the persistence of dominant
SDW correlations in the normal state. Antiferromagnetism becomes
unstable against the formation of a charge-density-wave state when $g_1^\perp$
exceeds some critical value. While these features persist when
both Umklapp processes and interchain forward scattering ($g_2^\perp$) are
taken into account, the effect of $g_2^\perp$ alone is found to frustrate
nearest-neighbor interchain $d$- and $f$-wave pairing and instead favor 
next-nearest-neighbor interchain singlet or triplet pairing. We argue 
that the close proximity of SDW and charge-density-wave phases,
singlet $d$-wave and triplet $f$-wave superconducting phases in the
theoretical phase
diagram provides a possible explanation for recent puzzling experimental 
findings in the Bechgaard salts, including the coexistence of
SDW and charge-density-wave 
phases and the possibility of a triplet pairing in the superconducting phase. 
\end{abstract}

\maketitle 

\section{\label{intro}Introduction} 

The theory of low-dimensional metals has exerted a 
strong influence on our understanding of 
ordered  phases in  quasi-one-dimensional (quasi-1D) organic conductors. 
The description of the extensively studied Bechgaard 
salts series ((TMTSF)$_2$X) and their sulfur analogs, 
the Fabre ((TMTTF)$_2$X) salts, has served over 
more than two decades to illustrate 
this view.\cite{Bourbon99,Jerome94,Dupuis05} 
For these materials, a direct correspondence 
can be traced between the various modulated spin and 
charge ordered states of their phase diagram and 
the possible states of the quasi-1D electron 
gas model when the couplings are repulsive and the filling 
of the band is commensurate with the underlying 
lattice.\cite{Emery82,Giamarchi97} In its standard form, this generic 
model is defined by the phenomenological expression 
of the direct interaction between electrons in terms of 
weak \textit{intra}chain backward ($g_1$) and forward ($g_2$) 
electron-electron scattering processes, to which 
Umklapp ($g_3$) scattering amplitudes are added at 
commensurate band filling.\cite{Dzyaloshinskii72,Barisic81,
Giamarchi97,Tsuchiizu01} The quasi-1D 
character of the model is defined by adding an interchain 
single electron hopping integral $t_\perp$, which is at 
least an order of magnitude smaller than its longitudinal 
counterpart.

In virtually all compounds of the above series, 
superconductivity is observed next to a 
spin-density-wave state for some critical value 
on the pressure scale,\cite{Jerome80,Balicas94,Adachi00,Jaccard01} 
whereas antiferromagnetic spin correlations are found to 
dominate the metallic state precursor to superconductivity 
over a wide range of temperatures.\cite{Bourbon84,Wzietek93} 
On the theoretical side, however, for spin-independent 
repulsive couplings and for a Fermi surface with good nesting 
properties, the coexistence of spin-density-wave (SDW)  
and superconducting (SC) correlations is essentially excluded 
from the model phase diagram.\cite{Emery79,Solyom79} 
It is only when deviations from  perfect electron-hole symmetry (nesting) 
are introduced and the long-range component of the 
SDW order is suppressed that 
superconductivity can be actually realized 
in place of magnetism.\cite{Emery86,Caron86} 
In the anisotropic metallic phase where this suppression 
takes place, interchain Cooper pairing is enhanced and 
superconductivity emerges from the coupling between the 
weakened electron-hole and the still singular 
electron-electron scattering channels.
Recent calculations \cite{Duprat01,Bourbon04,Fuseya05} 
using the renormalization group (RG) method did confirm 
the existence of such an electronic pairing mechanism 
beyond the level of single-channel RPA-like
approaches.\cite{Bealmonod86,
Scalapino86,Bourbon88,Shimahara89,Kino99} A smooth crossover 
from the SDW state to superconductivity 
has then been found whenever the amplitude of nesting 
deviations reaches some threshold -- a result 
in accordance with the sequence of transitions observed 
as a function of pressure.\cite{Jerome80,Wilhelm01,Adachi00} 

The RG approach shows nevertheless that whenever antiferromagnetism 
stands out as the dominant correlation in the normal state, 
the most stable interchain pairing is invariably a spin singlet 
state corresponding to a `$d$-wave' symmetry gap with nodes on the 
Fermi surface. Although some experimental findings in the 
Bechgaard salts and their sulfur analogs do agree with this 
type of pairing, \cite{Wilhelm01,Joo04} other series of 
observations have rather been interpreted in support of 
a triplet order parameter,\cite{Gorkov85,Lee97,Oh04,Lee02} 
thus challenging the singlet scenario for superconductivity. 
Though there is too little so far to favor one scenario 
over another, these observations bring us to the question of 
whether triplet superconductivity can be possible or not when 
short-range antiferromagnetic correlations are dominant in 
the metallic state. As is well known for the electron gas model, 
there is a region of the phase diagram where triplet 
`$p$-wave' superconductivity does exist as the most stable state. It has been
suggested, on a phenomenological basis, that such a state is realized in the
Bechgaard salts.\cite{Lebed00} Within a microscopic approach, the region 
where $p$-wave superconductivity is stable 
is defined by irrelevant Umklapp scattering 
and by a backward scattering coupling that is much larger than 
forward scattering. However, this description pattern 
for superconductivity can be considered unsatisfactory 
given the unrealistic constraint it puts on the coupling 
constants, and for the suppression of both the 
antiferromagnetism at short distance and the Mott pseudo gap 
in the charge sector.\cite{Dupuis05}
These flaws can hardly be reconciled 
with the related phenomenology of these molecular compounds 
observed around the critical pressure for superconductivity, 
when either the temperature or the magnetic field 
is varied.\cite{Bourbon99,Wzietek93,Vescoli98,Bourbon98,Chaikin96} 

Another way to look at this problem is to consider 
more closely the effect of charge fluctuations on 
superconductivity. Staggered charge fluctuations 
are known to favor triplet pairing at odd but large 
angular momentum, a mechanism whose roots go back 
to the early work of Kohn and Luttinger about 
Cooper pairing in the presence of charge -- Friedel -- oscillations 
in isotropic Fermi systems.\cite{Kohn65} For the 
quasi-1D electron gas model and 
its version for lattice electrons, recent calculations have 
shown that triplet $f$-wave pairing is indeed enhanced when 
intrachain couplings are chosen to boost 
charge-density-wave (CDW) fluctuations close to the level found in 
the SDW channel.\cite{Fuseya05,Kuroki01,Onari04,Tanaka04} 
However, for realistic repulsive couplings, $d$-wave pairing 
still remains tied to the highest critical temperature and 
hence to the most stable state for superconductivity. 

All this goes to establish the robustness of $d$-wave pairing 
for the model with repulsive intrachain couplings 
and nesting deviations. It turns out, however, that the 
model is incomplete when charge fluctuations are found 
along the chains since then \textit{inter}chain Coulomb 
interaction is also present in practice. 
The inclusion of direct interchain electron-electron 
scattering processes, which will be denoted by 
$g_{i=1,2,3}^\perp$ in the following, defines the 
quasi-1D electron gas model in its 
extended form.\cite{Gorkov74} At large momentum transfer, 
the interchain interaction is well known to favor a 
CDW ordered 
state.\cite{Gorkov74,Mihaly76,Menyhard77,Lee77,
Saub76} 
This mechanism 
is mostly responsible for CDW long-range 
order observed in several organic and inorganic 
low-dimensional solids.\cite{Pouget89,Barisic85} 
The physical relevance of interchain interactions 
in the Bechgaard salts, besides the intrachain $g_i$ and $t_\perp$,
is supported by x-ray studies, which revealed that the 
SDW phase of these compounds is actually 
accompanied by CDW order.\cite{Pouget96,Pouget97,Kagoshima99} 
On the theoretical 
grounds, very little is known about the impact of adding 
direct interchain interactions on the structure 
of the phase diagram, especially in the repulsive 
sector when both a finite $t_\perp$ and 
nesting deviations are present. 

In this work we wish to determine the possible 
density-wave and superconducting states of the 
extended quasi-1D electron-gas model. 
In order to tackle this problem we shall apply 
the renormalization group method, which at 
the one-loop level has proved to be suited to reach 
a controlled description of interfering 
density-wave and superconducting channels of correlations. 
Among the results reported below, we have the unexpected 
finding that a small repulsive interchain backscattering 
term $g_1^\perp>0$ is sufficient to make $d$-wave 
superconductivity unstable to the benefit of a triplet $f$-wave 
phase. This occurs despite dominant SDW 
correlations in the metallic state and stable itinerant 
antiferromagnetism at lower nesting deviations. Under 
the latter conditions, SDW order becomes in turn 
unstable to the formation of a CDW state 
when the amplitude of $g_1^\perp$ exceeds some critical 
value. While these features persist when commensurability 
effects are taken into account and 
small -- half-filling -- Umklapp scattering is included, 
the effect of interchain forward 
scattering \hbox{($g_2^\perp>0$)} is found to frustrate 
nearest-neighbor interchain $d$- and 
$f$-wave pairing and to favor instead  
superconductivity with next-nearest-neighbor interchain 
pairing. Part of these results have been reported in
Refs.~\onlinecite{Bourbon04b,Nickel05}.  

In Section~\ref{technique}, we introduce the model
and the RG scheme employed for the four point vertices and 
the response functions. By way of illustration, the RG results 
at the one-loop level are given for purely intrachain
interactions. In Section~\ref{innter}, we 
present the results for non zero interchain backward 
and forward interactions by which the different 
possibilities of ordered states in the phase diagram are 
obtained in the incommensurate case. The influence of Umklapp 
processes in the half-filled case is examined
in Section~\ref{umklapp}. A discussion of the results is given 
in section~\ref{nnandexp}, where a possible connection between 
theory and experiments is made.

\section{The extended electron gas model}
\label{technique}

\subsection{Model}
\label{model}
We consider a lattice of $N_\perp$ coupled metallic chains described 
by the partition function 
$Z = \int\!\!\int D\psi^*D\psi\, e^{-(S_0 + S_I)}$, 
which is  expressed as a functional integration over anticommuting 
(Grassmann) fields $\psi$. Here $S_0$ and $S_I$ are the 
noninteracting and interacting parts of the action, respectively.   
The former part is given by 
\begin{equation} 
S_0 = - \sum_K \psi_K\konjug (ik_0-\xi_{\bf k}) \psi_K \ , 
\end{equation} 
where $K=(\sigma,k)$, $k=(k_0;\mathbf{k})$, $k_0$ is the fermionic Matsubara 
frequency,   
$\mathbf{k}=(k\upara,k\uperp)$, the wave vector, and $\sigma=\pm$, the
spin of the fermion field. The 
kinetic energy is given by 
\begin{eqnarray} 
\label{spectrum} 
\xi_{\mathbf{k}} &=& \epsilon_{\mathbf{k}}-\mu = 
v_F ( \betrag{k\upara} - k_F ) + 
\epsilon\uperp(k\uperp) \ , \\ 
\label{perpkin} 
\epsilon\uperp(k\uperp) &=& 
-2t\uperp\cos k\uperp - 2t\uperp'\cos 2k\uperp \ , 
\end{eqnarray} 
where $\mu$ is the chemical potential  taken as temperature 
independent. Here we have linearized the spectrum in the chain direction, 
using $v_F=2 t\upara \sin k_F$ as the longitudinal Fermi velocity, 
$k_F$ being the parallel Fermi wave vector if $t\uperp = 0$. 
Throughout this work both the chain lattice constant 
$a$ and the interchain distance $d_\perp$ are put equal to unity. 
The  interchain single electron hopping $t_\perp$  is considered 
small with respect to the longitudinal bandwidth $2\Lambda_0$. 
In the following, we will take  $2\Lambda_0=30 t_\perp$, which is a 
typical figure for the anisotropy ratio in quasi-1D 
conductors like the Bechgaard and 
Fabre salts.\cite{Grant83,Ducasse86} 
The next nearest-neighbor hopping $t\uperp'\ll t_\perp$  
in the transverse direction gives the amplitude of nesting deviations.
In Eq.~(\ref{perpkin}), we have neglected the possibility of hopping
in the third direction, which does not have any sizeable effect on
our calculations. Its existence is of course crucial for the
stabilization of true long-range order at finite temperature. 

In the framework of the quasi-1D electron gas 
model,\cite{Dzyaloshinskii72,Solyom79} 
the electron-electron interaction is parameterized by means of the g-ology 
approach. One first distinguishes between right and left moving fermions, 
depending on their velocity along the chains, 
so that the Grassmann variables $\psi_K^{(*)}$ become  
\begin{equation} 
\psi_K^{(*)} = 
\begin{cases} 
R_K^{(*)} & \text{ if } k\upara>0 \ , \\ 
L_K^{(*)} & \text{ if } k\upara<0 \ . 
\end{cases} 
\end{equation} 
Using these definitions, the interaction part of the action 
takes the form 
\begin{eqnarray} 
S_I &=& \frac{T}{N} \sum_{k_1'k_2'k_2k_1} \sum_{\sigma\sigma'} 
\delta_{k_1'+k_2',k_2+k_1\ {\rm mod}\ G} \\ 
&& \nonumber 
\times \Bigl\{ \ww_1(k_1'k_2'k_2k_1) 
R^*_{k_1'\sigma} L^*_{k_2'\sigma'} 
R_{k_2\sigma'} L_{k_1\sigma} \\ 
&& \nonumber
+ \ww_2(k_1'k_2'k_2k_1) 
R^*_{k_1'\sigma} L^*_{k_2'\sigma'} 
L_{k_2\sigma'} R_{k_1\sigma} \\ 
&& \nonumber
+ \frac{1}{2} \bigl( \ww_3(k_1'k_2'k_2k_1) 
R^*_{k_1'\sigma} R^*_{k_2'\sigma'} 
L_{k_2\sigma'} L_{k_1\sigma} + {\rm c.c.} \bigr) \Bigr\} \ ,  
\end{eqnarray} 
where $N$ is the number of lattice sites and $G = (0;4k_F,0)$ 
the reciprocal lattice vector along the chains, which is involved 
in the Umklapp scattering at half-filling. 
Here we have omitted the so-called $g_4$ contribution for the  
interaction of electrons of the same branch since this coupling 
does not contribute to the singular channels of correlation in the   
renormalization group flow at the one-loop level. If we restrict 
ourselves to intrachain and nearest-neighbor chain
interactions, the amplitudes of the bare interactions are given by 
\begin{equation} 
\label{gstarting} 
g_j(k_{\bot 1}' k_{\bot 2}' k_{\bot 2} k_{\bot 1}) =
g_{j} + 2 \gnn_j \cos (k_{\bot 1}'-k_{\bot 1}) \ ,
\end{equation} 
where the only -- transverse -- momentum dependence comes from the
interchain interaction. In this work  
we analyze the properties of the model for repulsive 
$g_{j}$, $g_j^\perp$ and finite Umklapp scattering, which is  
the physically relevant sector for real materials like the 
Bechgaard salts and their sulfur analogs.

\subsection{Renormalization group equations 
for the interactions} 
\label{renorm}
The renormalization group can be applied to  
the problem of quasi-1D systems of  
interacting electrons in different 
ways.\cite{Duprat01,Bourbon02,Honerkamp03,Dusuel03} 
In the following, we shall adopt the so-called 
one-particle irreducible (1PI) momentum 
shell RG scheme, 
as developed by Honerkamp \textit{et al.}\cite{Honerkamp01b} (cf.\ also 
Binz \textit{et al.}\cite{Binz03}). In this scheme, the 1PI vertex 
functions $\vertex_{\Lambda}(K_j)$ of a physical system 
with infrared cutoff $\Lambda$ are calculated. 
Only degrees of freedom with energies equal to or
greater than the cutoff are integrated out, which allows one to establish a
direct link with the low energy effective action $S_{\Lambda}^{eff}$ with an
\textit{ultraviolet} cutoff, as obtained in a Kadanoff-Wilson
renormalization group scheme. According to 
Morris,\cite{Morris94} if $\betrag{\epsilon_{{\bf k}_j}}<\Lambda$, the 
effective interactions $g_{\Lambda}^{eff}(K_j)$ that appear  in 
$S_{\Lambda}^{eff}$ are simply 
$g_{\Lambda}^{eff}(K_j)=\vertex_{\Lambda}(K_j)$. 

The free particle propagator is suppressed for energies 
below an infrared cutoff $\Lambda$, 
\begin{equation} 
C_K = \mittelwert{\psi_K \psi_K\konjug}_{S_0} = 
\Theta(\betrag{\xi_{\bf k}}-\Lambda) \times 
\frac{-1}{ik_0-\xi_{\bf k}} \ , 
\end{equation} 
where $\Theta$ is the step function. 
Setting all vertices involving more than two 
particles to zero, the renormalization group 
equations to the one-loop level are obtained. 
We also neglect self-energy corrections. 

It is practically impossible to take into account the complete 
functional dependence of the vertices, which will in general vary 
with the wave vectors ${\bf k}$ as well as with the frequencies 
$k_0$ of the incoming and outgoing particles. By means of scaling 
arguments, one can show that the frequency dependence is irrelevant 
in the renormalization group sense, cf.\ for 
example Ref.~\onlinecite{Shankar94}. 
The same is true for the dependence on the \textit{distance} 
of the wave vector ${\bf k}$ from the Fermi surface. 
Irrelevant variables are not necessarily negligible, in the sense that
they may in principle have an influence on the RG flow before they
vanish. However, the ${\bf k}_s$ dependence, with ${\bf k}_s$ the
projection of the wave vector on the Fermi surface, 
remains by far the most important one in the low energy 
limit. We will hence only consider the dependence of the functions 
$\vertex$ on the positions of the wave vectors along the Fermi surface, 
which we will parameterize by $k\uperp$. 
Due to momentum conservation, it is in general impossible to have all
four arguments of a given vertex on the Fermi surface. Usually, the
flow of the vertices with three arguments on the Fermi 
surface is calculated. In the case of imperfect nesting, 
such a procedure will underestimate the 
SDW correlations. In our case, 
the strong anisotropy of the dispersion relation allows us to make a
slightly different choice. We fix the values 
of $k\upara$ and $k_0$ such that the most divergent contributions 
to the renormalization group flow are 
taken into account. 
If we define the Cooper, Landau, and Peierls momentum variables 
\begin{eqnarray} 
\label{qchanneldefs}
q_{\Cooper} & = & k_1+k_2 \ , \\ 
\nonumber 
q_{\Landau} & = & 
\begin{cases} 
k_1'-k_2 & \text{ for } \vertex_1 \ , \\ 
k_1'-k_1 & \text{ for } \vertex_2 \ , 
\end{cases} \\ 
\nonumber 
q_{\Peierls} & = & 
\begin{cases} 
k_1'-k_1 & \text{ for } \vertex_1 \ , \\ 
k_1'-k_2 & \text{ for } \vertex_2 \text{ and } \vertex_3 \ , 
\end{cases} \\ 
\nonumber 
q_{\Peierls}' & = & k_1'-k_1 \text{ for } \vertex_3 \ , 
\end{eqnarray} 
we can write the one-loop RG equations in the form 
\begin{eqnarray*} 
\frac{d}{dl} \vertex(q_{\Cooper},q_{\Landau},q_{\Peierls})
& = & \sum_{k_{\rm loop}} \text{ Cooper}
\bigl(\vertex(q_{\Cooper},\tilde{q}_{\Landau},\tilde{q}_{\Peierls})\bigr) \\ 
&& + \sum_{k_{\rm loop}} \text{ Peierls}
\bigl(\vertex(\tilde{q}_{\Cooper},\tilde{q}_{\Landau},q_{\Peierls})\bigr) \
. 
\end{eqnarray*}
In this equation, $\tilde q_{\Cooper}$, 
$\tilde q_{\Landau}$ and $\tilde q_{\Peierls}$ are functions 
of the internal loop momentum-frequency $k_{\rm loop}$. The most
divergent contributions to the RG flow are obtained if 
$q_{\Cooper,0} = q_{\Peierls,0}^{(\prime)} = 0$ for the 
frequencies, as well as 
\begin{eqnarray}
\label{leadingqpara}
\qcooperpara & = & 0 \ , \; q_{\Peierls\|} \ = \ 2k_F
\text{ for } \vertex_{1,2} \ , \\ 
\nonumber 
q_{\Peierls\|}^{(\prime)} & = & \pm 2k_F
\text{ for } \vertex_{3} \ , 
\end{eqnarray} 
for the longitudinal components. 
Note that, for either choice of the $k\upara$ dependence, one does 
not obtain a closed set of equations. The renormalization group
equations for vertices for which relations \glbezug{leadingqpara} 
are valid contain vertex functions with \textit{different} values of 
$q_{\Cooper}$, $q_{\Peierls}$. We will therefore 
need to replace these functions by those 
calculated by imposing conditions~(\ref{leadingqpara}). 

For the dimensionless vertices $\tilde\Gamma=\Gamma/\pi v_F$, 
we thus obtain the equations\cite{Nickel04} 
\begin{widetext}
\begin{eqnarray} 
\dot{\tilde{\vertex}}_1(k_{\bot 1}'k_{\bot 2}'k_{\bot 2}k_{\bot 1}) & = & 
- \frac{1}{N\uperp} \sum_{k\uperp} \biggl[ 
B_{\Cooper}(k\uperp,\qcooperperp) \\ 
\nonumber && \times \Bigl\{ 
\tilde{\vertex}_1(k_{\bot 1}'k_{\bot 2}'k\uperp k\uperp') 
\tilde{\vertex}_2(k\uperp k\uperp'k_{\bot 1}k_{\bot 2}) \\ 
\nonumber && + \tilde{\vertex}_2(k_{\bot 1}'k_{\bot 2}'k\uperp'k\uperp) 
\tilde{\vertex}_1(k\uperp k\uperp'k_{\bot 2}k_{\bot 1}) 
\Bigr\} 
\biggr] \biggr|_{\zweikleinemathezeilen{
k\uperp'=-k\uperp+\qcooperperp}{
\qcooperperp=k_{\bot 1}+k_{\bot 2}} } 
\\
\nonumber && + 
\frac{1}{N\uperp} \sum_{k\uperp} \biggl[ 
B_{\Peierls}(k\uperp,\qpeierlsperp) \\ 
\nonumber && \times 
\Bigl\{ 
\bigl( \tilde{\vertex}_2(k_{\bot 1}'k\uperp k_{\bot 1} k\uperp') 
- \tilde{\vertex}_1(k_{\bot 1}'k\uperp k\uperp'k_{\bot 1}) \bigr) 
\tilde{\vertex}_1(k\uperp'k_{\bot 2}'k_{\bot 2} k\uperp) \\ 
\nonumber && + \tilde{\vertex}_1(k_{\bot 1}'k\uperp k\uperp'k_{\bot 1}) 
\bigl( \tilde{\vertex}_2(k\uperp'k_{\bot 2}'k\uperp k_{\bot 2}) 
- \tilde{\vertex}_1(k\uperp'k_{\bot 2}'k_{\bot 2} k\uperp) \bigr) \\ 
\nonumber && + \bigl( 
\tilde{\vertex}_3(k_{\bot 1}'k\uperp k_{\bot 1} k\uperp') 
- \tilde{\vertex}_3(k_{\bot 1}'k\uperp k\uperp'k_{\bot 1}) \bigr) 
\tilde{\vertex}_3(k\uperp'k_{\bot 2}'k_{\bot 2} k\uperp) \\ 
\nonumber && + \tilde{\vertex}_3(k_{\bot 1}'k\uperp k\uperp'k_{\bot 1}) 
\bigl( \tilde{\vertex}_3(k\uperp'k_{\bot 2}'k\uperp k_{\bot 2}) 
- \tilde{\vertex}_3(k\uperp'k_{\bot 2}'k_{\bot 2} k\uperp) \bigr) 
\Bigr\} 
\biggr] \biggr|_{\zweikleinemathezeilen{ 
k\uperp'=k\uperp+\qpeierlsperp}{ 
\qpeierlsperp=k_{\bot 1}'-k_{\bot 1}} } \ , 
\end{eqnarray} 
\begin{eqnarray} 
\dot{\tilde{\vertex}}_2(k_{\bot 1}'k_{\bot 2}'k_{\bot 2}k_{\bot 1}) & = & 
- \frac{1}{N\uperp} \sum_{k\uperp} \biggl[ 
B_{\Cooper}(k\uperp,\qcooperperp) \\ 
\nonumber && \times 
\Bigl\{ 
\tilde{\vertex}_1(k_{\bot 1}'k_{\bot 2}'k\uperp k\uperp') 
\tilde{\vertex}_1(k\uperp k\uperp'k_{\bot 1}k_{\bot 2}) \\ 
\nonumber && + \tilde{\vertex}_2(k_{\bot 1}'k_{\bot 2}'k\uperp'k\uperp) 
\tilde{\vertex}_2(k\uperp k\uperp'k_{\bot 2}k_{\bot 1}) 
\Bigr\} 
\biggr] \biggr|_{\zweikleinemathezeilen{
k\uperp'=-k\uperp+\qcooperperp}{
\qcooperperp=k_{\bot 1}+k_{\bot 2}} } 
\\ 
\nonumber && + 
\frac{1}{N\uperp} \sum_{k\uperp} \biggl[ 
B_{\Peierls}(k\uperp,\qpeierlsperp) \\ 
\nonumber && \times 
\Bigl\{ 
\tilde{\vertex}_2(k_{\bot 1}'k\uperp k_{\bot 2} k\uperp') 
\tilde{\vertex}_2(k\uperp'k_{\bot 2}'k\uperp k_{\bot 1}) \\ 
\nonumber && + \tilde{\vertex}_3(k_{\bot 1}'k\uperp k_{\bot 2} k\uperp') 
\tilde{\vertex}_3(k\uperp'k_{\bot 2}'k\uperp k_{\bot 1}) 
\Bigr\} 
\biggr] \biggr|_{\zweikleinemathezeilen{ 
k\uperp'=k\uperp+\qpeierlsperp}{ 
\qpeierlsperp=k_{\bot 1}'-k_{\bot 2}} } \ ,
\end{eqnarray} 
\begin{eqnarray} 
\dot{\tilde{\vertex}}_3(k_{\bot 1}'k_{\bot 2}'k_{\bot 2}k_{\bot 1}) & = & 
\frac{1}{N\uperp} \sum_{k\uperp} 
\biggl[ B_{\Peierls}(k\uperp,\qpeierlsperp') \\ 
\nonumber && \times 
\Bigl\{ 
\bigl( \tilde{\vertex}_2(k_{\bot 1}'k\uperp k_{\bot 1} k\uperp') 
- \tilde{\vertex}_1(k_{\bot 1}'k\uperp k\uperp' k_{\bot 1}) \bigr) 
\tilde{\vertex}_3(k\uperp' k_{\bot 2}' k_{\bot 2} k\uperp) \\ 
\nonumber && + \tilde{\vertex}_1(k_{\bot 1}'k\uperp k\uperp' k_{\bot 1}) 
\bigl( \tilde{\vertex}_3(k\uperp' k_{\bot 2}' k\uperp k_{\bot 2}) 
- \tilde{\vertex}_3(k\uperp' k_{\bot 2}' k_{\bot 2} k\uperp) \bigr) \\ 
\nonumber && + \bigl( 
\tilde{\vertex}_3(k_{\bot 1}'k\uperp k_{\bot 1} k\uperp') 
- \tilde{\vertex}_3(k_{\bot 1}'k\uperp k\uperp' k_{\bot 1}) \bigr) 
\tilde{\vertex}_1(k\uperp' k_{\bot 2}' k_{\bot 2} k\uperp) \\ 
\nonumber && + \tilde{\vertex}_3(k_{\bot 1}'k\uperp k\uperp' k_{\bot 1}) 
\bigl( \tilde{\vertex}_2(k\uperp' k_{\bot 2}' k\uperp k_{\bot 2}) 
- \tilde{\vertex}_1(k\uperp' k_{\bot 2}' k_{\bot 2} k\uperp) \bigr) 
\Bigr\} 
\biggr] \biggr|_{ \zweikleinemathezeilen{ 
k\uperp'=k\uperp+\qpeierlsperp'}{ 
\qpeierlsperp' = k_{\bot 1}'-k_{\bot 1} } } 
\\ 
\nonumber && + \frac{1}{N\uperp} \sum_{k\uperp} \biggl[ 
B_{\Peierls}(k\uperp,\qpeierlsperp) \\ 
\nonumber && \times 
\Bigl\{ 
\tilde{\vertex}_2(k_{\bot 1}'k\uperp k_{\bot 2} k\uperp') 
\tilde{\vertex}_3(k\uperp'k_{\bot 2}'k\uperp k_{\bot 1}) \\ 
\nonumber && + \tilde{\vertex}_3(k_{\bot 1}'k\uperp k_{\bot 2} k\uperp') 
\tilde{\vertex}_2(k\uperp' k_{\bot 2}' k\uperp k_{\bot 1}) 
\Bigr\} 
\biggr] \biggr|_{\zweikleinemathezeilen{ 
k\uperp'=k\uperp+\qpeierlsperp}{ 
\qpeierlsperp=k_{\bot 1}'-k_{\bot 2}} } \ , 
\end{eqnarray} 
\end{widetext} 
where the dot denotes derivation with respect to 
$-\ln(\Lambda/\Lambda_0)$. 
The particle-particle ($\Cooper$) and particle-hole ($\Peierls$)
loops, after summation over Matsubara frequencies and longitudinal
wave vectors, are given by 
\begin{eqnarray} 
B_{\Cooper/\Peierls}(k\uperp,q\uperp) & =& 
\sum_{\nu=\pm 1} 
\Theta ( \betrag{\Lambda + \nu A_{\Cooper/\Peierls} } - \Lambda ) \\ 
\nonumber && \times \frac{1}{2} \left( 
\tanh \frac{\Lambda+ \nu A_{\Cooper/\Peierls}}{2T} 
+ \tanh \frac{\Lambda}{2T} \right) \\ 
\nonumber && \times \frac{\Lambda}{2\Lambda+\nu A_{\Cooper/\Peierls}} \ , 
\end{eqnarray} 
\begin{eqnarray} 
A_{\Cooper}(k\uperp,q\uperp) & = & -\epsilon\uperp(k\uperp) 
+\epsilon\uperp(-k\uperp+q\uperp) \ , \\ 
\nonumber 
A_{\Peierls}(k\uperp,q\uperp) & = & -\epsilon\uperp(k\uperp) 
-\epsilon\uperp(k\uperp+q\uperp) \ . 
\end{eqnarray} 
For continuity reasons, we always take $\Theta(0) := \frac{1}{2}$. 
The starting values are given by the bare 
interactions, cf.\ section~\ref{model}. 

It is useful to consider the following limiting cases for the RG
equations. As long as $\Lambda \gg t\uperp$, we may take their
1D limit ($t\uperp=0$). In this regime, the Cooper and
Peierls renormalization channels are
entirely coupled, in the sense that all vertices are strongly 
renormalized in both channels (except for Umklapp processes, 
which only appear in the Peierls channel). 
When $\Lambda \ll t\uperp$, the coupling between the Cooper
and Peierls channels is weak in the sense that, depending on the arguments
$k_\perp$, most vertices are strongly renormalized in only one (or no)
channel at a time.
Nevertheless, the remaining interplay between the channels 
is at the origin of spin- and charge-fluctuation-induced
superconductivity in the weak coupling regime. 
For purely repulsive interactions, the Peierls
channel is the most important one as long as deviations from perfect
nesting may be neglected, \textit{i.~e.\ }for $\Lambda \gg t\uperp'$. 
In the presence of attractive effective 
interactions, the Cooper channel plays
an important role, and will be dominant when $\Lambda \ll
t\uperp'$. The RG equations are written in these limits in
appendix~\ref{glims}. 

A similar dimensional crossover may be observed with respect to the 
temperature instead of $\Lambda$. As long as $T \gg t_\perp$, 
the functions we calculate do not vary very strongly with 
the transverse wave vector,
contrary to the low temperature case. 
An example is given in Fig.~\ref{chiTs}(a).

\subsection{\label{resprenorm}Response functions} 

In order to evaluate susceptibilities, we add a term of the form 
\begin{eqnarray} 
S_h & = & \sum_{\alpha} \sum_q h_{\alpha SC}^*(q) 
O_{\alpha SC}(q) + {\rm c.c.} \\ 
\nonumber && + \sum_{\alpha,M} \sum_q h_{\alpha DW}^{(M)*}(q) 
O_{\alpha DW}^{(M)}(q) + {\rm c.c.} 
\end{eqnarray} 
to the action. The first term describes pairing 
and the second term density-wave correlations. 
The external fields $h^{(*)}(q)$ are 
taken to be infinitesimal. They couple to pairs of fermionic
variables which we will define in the following. Let us first
introduce the particle-particle operators 
\begin{equation} 
o_{\alpha}(k,q) = \sum_{\sigma'\sigma} 
\tau^{(\alpha)}_{\sigma'\sigma} 
L_{-k+q,\sigma'} R_{k,\sigma} 
\end{equation} 
for singlet ($\alpha=s$) and triplet ($\alpha=t_{x,y,z}$) pairs. 
The spin dependence is given by the coefficients 
\begin{eqnarray} 
&& \tau^{(s)}_{\sigma'\sigma} = \sigma \delta_{\sigma',-\sigma} \ , \;
\\ 
\nonumber && 
\tau^{(t_x)}_{\sigma'\sigma} = -\sigma \delta_{\sigma'\sigma} \ , \; 
\tau^{(t_y)}_{\sigma'\sigma} = -i \delta_{\sigma'\sigma} \ , \;
\tau^{(t_z)}_{\sigma'\sigma} = \delta_{\sigma',-\sigma} 
\end{eqnarray} 
($\sigma=\pm 1$). The pair operator appearing in $S_h$ is defined
as 
\begin{equation} 
O_{\alpha SC}(q) = \sqrt{\frac{T}{N}} 
\sum_{ \{ k; k\upara > 0 \} } z_{\alpha}(q-k,k) o_{\alpha}(k,q) \ . 
\end{equation} 
The function $z_{\alpha}$ 
describes the orbital symmetry of the superconducting
order parameter. We classify these order parameters 
by their behavior on the Fermi surface. They are
parameterized by $k\uperp$ and $r=\signum\ k\upara$. 
A list of the superconducting order parameters we shall examine is 
given in Table~\ref{opetab}. With these, we have 
\begin{equation} 
z_{\alpha}(k\uperp',k\uperp) = 
\begin{cases} 
1 & \text{ for $s$ and $p_x$, } \\ 
\sqrt{2} \Delta_{r=+}(k_\perp) & \text{ for all others.} 
\end{cases} 
\end{equation} 

\begin{table} 
\[ 
\begin{array}{|l|l|l|}
\hline
\text{name } & \text{spin pairing} & \Delta_r(k_\perp) 
\\ \hline \hline 
\begin{array}{l} s \end{array} & \text{singlet } 
& \begin{array}{l} 1 \end{array} 
\\ \hline 
\begin{array}{l} p_x \\ p_y \end{array}
& \text{triplet } 
& \begin{array}{l} r \\ \sin k\uperp \end{array} 
\\ \hline 
\begin{array}{l} d_{x^2-y^2} \\ d_{xy} \end{array} 
& \text{singlet } 
& \begin{array}{l} \cos k\uperp \\ r\ \sin k\uperp \end{array}
\\ \hline 
\begin{array}{l} f \end{array} 
& \text{triplet } 
& \begin{array}{l} r\ \cos k\uperp \\ \sin 2k\uperp \end{array}
\\ \hline 
\begin{array}{l} g \end{array} 
& \text{singlet } 
& \begin{array}{l} \cos 2k\uperp \\ r \sin 2k\uperp \end{array}
\\ \hline
\begin{array}{l} h \end{array} 
& \text{triplet } 
& \begin{array}{l} r \cos 2k\uperp \end{array} 
\\ \hline 
\begin{array}{l} i \end{array} 
& \text{singlet } 
& \begin{array}{l} \cos 3k\uperp \end{array} 
\\ \hline 
\end{array}
\] 
\caption{\label{opetab}Superconducting order parameters in 
a quasi-1D geometry. 
The names are assigned according to the number of sign changes 
along the Fermi surface.} 
\end{table} 

Similarly, we introduce particle-hole operators 
\begin{equation} 
o_{\alpha}(k,q) = \sum_{\sigma'\sigma} 
\sigma^{(\alpha)}_{\sigma'\sigma} 
L^*_{k-q,\sigma'} R_{k,\sigma} 
\end{equation} 
for charge ($\alpha=C$) and spin ($\alpha=S_{x,y,z}$)
excitations. Here, $\sigma^{(C)}$ is the $2 \times 2$ identity matrix,
and $\sigma^{(S_{x,y,z})}$ are the Pauli matrices. 
At half filling, one has to distinguish between bond and site density
waves. In direct space, these are given by 
\begin{equation} 
O_x = 
\begin{cases} 
\frac{1}{2} \sum_{\sigma'\sigma} \sigma^{(\alpha)}_{\sigma'\sigma} 
\psi_{x,\sigma'}^* \psi_{x,\sigma} & \text{ for site DW, }\\ 
\frac{1}{4} \sum_{\sigma'\sigma} \bigl( 
\sigma^{(\alpha)}_{\sigma'\sigma} 
\psi_{x,\sigma'}^* \psi_{x+d,\sigma} 
+ c.c. \bigr) & \text{ for bond DW, }
\end{cases} 
\end{equation} 
where $d=(0;d\upara,0)$, and $d\upara$ is the 
lattice periodicity along the chains. Since we have 
$k_F d\upara =\frac{\pi}{2}$, the 
associated Fourier transforms are, for $q\upara \approx 2k_F$, 
\begin{widetext} 
\begin{equation} 
\label{odsitelien} 
O_{\alpha DW}^{(M)}(q) \approx \frac{1}{2} \sqrt{\frac{T}{N}} 
\sum_{ \{ k; k\upara>0 \} } \bigl[ 
z_{\alpha}^{(M)}(k-q,k) o_{\alpha}(k,q) 
+ M \bigl( z_{\alpha}^{(M)}(k+q-G,k) \bigr)\konjug 
o\konjug_{\alpha}(k,G-q) \bigr] 
\end{equation} 
\end{widetext} 
with $M=+$ for site and $M=-$ for bond density waves, and 
$z_{\alpha}^{(M)}(k,q)=1$. (For the case $M=-1$, we have 
neglected a constant imaginary factor.) Note that 
$(O_{\alpha}^{(M)})^*=MO_{\alpha}^{(M)}$, and that 
$\mittelwert{O_{\alpha}^{(M)}O_{\alpha}^{(M)}}$ and 
$\mittelwert{O_{\alpha}^{(M)*}O_{\alpha}^{(M)*}}$ do 
not vanish, but contribute to the associated susceptibilities. 
Away from half filling, we only consider site density waves. 
In this case, 
only the first term in Eq.~(\ref{odsitelien}) contributes. 

In the presence of $S_h$, the one-particle vertex contains 
a nondiagonal part with a contribution 
linear in $(h_{\alpha DW}^{(M)}(q))^*$ of the form 
\begin{equation} 
\sigma^{(\alpha)}_{\sigma'\sigma} \delta_{k',k-q} \ 
\orbvertpart_{\alpha}^{(M)}(k',k) 
\end{equation} 
(if $k\upara > 0$). Also due to $S_h$, 
vertices with two outgoing, but no incoming, 
particles are now non-zero. Their linear part in 
$(h_{\alpha SC}(q))^*$ takes the form 
\begin{equation} 
\tau^{(\alpha)}_{\sigma'\sigma} \delta_{k',q-k} \ 
\orbvertpart_{\alpha}(k',k) 
\end{equation} 
(if $k\upara > 0$). The vertex parts 
$\orbvertpart_{\alpha}$ determine the renormalization of
the susceptibilities $\chi_{\alpha}$ as follows. In the beginning of
the flow, we have $\orbvertpart_{\alpha}(k',k)=z_{\alpha}(k',k)$ and 
$\chi_{\alpha}=0$. The RG equations for 
density waves ($\alpha=C,S$) are: 
\begin{widetext} 
\begin{eqnarray}  
\label{zDWrenorm} 
\dot{\orbvertpart}_{\alpha}^M (p\uperp,p\uperp+q\uperp) & = & 
\frac{1}{N\uperp} \sum_{k\uperp} 
B_{\Peierls}(k\uperp,q\uperp) 
\orbvertpart_{\alpha}^M (k\uperp,k\uperp+q\uperp) 
\\ 
\nonumber && \times \bigl[ 
\tilde{\vertex}_{\alpha}(k\uperp+q\uperp,p\uperp,k\uperp,p\uperp+q\uperp)
\\ 
\nonumber && + \delta_{q\upara,2k_F^{1D}} 
(\delta_{q\uperp,0} + \delta_{q\uperp,\pi/b}) \ M \ 
\tilde{\vertex}_3^{(\alpha)}(k\uperp,p\uperp,k\uperp+q\uperp,p\uperp+q\uperp) 
\bigr] \ , \\ 
\dot{\tilde{\chi}}_{\alpha}^M (q\uperp) 
& = & - \frac{2}{N\uperp} \sum_{k\uperp}
B_{\Peierls}(k\uperp,q\uperp) 
\betrag{\orbvertpart_{\alpha}^M (k\uperp,k\uperp+q\uperp)}^2 \ .
\end{eqnarray} 
For Cooper pairs ($\alpha=s,t$), we have: 
\begin{eqnarray} 
\label{zstrenorm} 
\dot{\orbvertpart}_{\alpha} (-p\uperp+q\uperp,p\uperp) & = & 
\frac{1}{N\uperp} \sum_{k\uperp} 
B_{\Cooper}(k\uperp,q\uperp) 
\orbvertpart_{\alpha}(-k\uperp+q\uperp,k\uperp) \\
\nonumber && \times 
\tilde{\vertex}_{\alpha}(k\uperp,-k\uperp+q\uperp,-p\uperp+q\uperp,p\uperp) 
\ , \\ 
\dot{\tilde{\chi}}_{\alpha} (q\uperp) 
& = & - \frac{2}{N\uperp} \sum_{k\uperp} 
B_{\Cooper}(k\uperp,q\uperp)
\betrag{\orbvertpart_{\alpha}(-k\uperp+q\uperp,k\uperp)}^2 \ . 
\end{eqnarray} 
$\tilde{\chi}$ is defined as $\pi v_{F\|} \chi$. 
We have introduced the linear combinations of vertex functions 
\begin{eqnarray} 
\label{CSstvertexdef}
\vertex_C(k_{\bot 1}', k_{\bot 2}', k_{\bot 2}, k_{\bot 1}) &=& 
-2 \vertex_1(k_{\bot 1}', k_{\bot 2}', k_{\bot 1}, k_{\bot 2}) 
+ \vertex_2(k_{\bot 1}', k_{\bot 2}', k_{\bot 2}, k_{\bot 1}) \ , \\ 
\nonumber 
\vertex_S(k_{\bot 1}', k_{\bot 2}', k_{\bot 2}, k_{\bot 1}) &= &
\vertex_2(k_{\bot 1}', k_{\bot 2}', k_{\bot 2}, k_{\bot 1}) \ , \\ 
\nonumber 
\vertex_3^{(C)}(k_{\bot 1}', k_{\bot 2}', k_{\bot 2}, k_{\bot 1}) &= &
-2 \vertex_3(k_{\bot 1}', k_{\bot 2}', k_{\bot 1}, k_{\bot 2}) 
+ \vertex_3(k_{\bot 1}', k_{\bot 2}', k_{\bot 2}, k_{\bot 1}) \ , \\ 
\nonumber 
\vertex_3^{(S)}(k_{\bot 1}', k_{\bot 2}', k_{\bot 2}, k_{\bot 1}) &= &
\vertex_3(k_{\bot 1}', k_{\bot 2}', k_{\bot 2}, k_{\bot 1}) \ , \\ 
\nonumber 
\vertex_s(k_{\bot 1}', k_{\bot 2}', k_{\bot 2}, k_{\bot 1}) &= &
- \vertex_1(k_{\bot 1}', k_{\bot 2}', k_{\bot 1}, k_{\bot 2}) 
- \vertex_2(k_{\bot 1}', k_{\bot 2}', k_{\bot 2}, k_{\bot 1}) \ , \\ 
\nonumber 
\vertex_t(k_{\bot 1}', k_{\bot 2}', k_{\bot 2}, k_{\bot 1}) &= &
\vertex_1(k_{\bot 1}', k_{\bot 2}', k_{\bot 1}, k_{\bot 2}) 
- \vertex_2(k_{\bot 1}', k_{\bot 2}', k_{\bot 2}, k_{\bot 1}) \ . 
\end{eqnarray} 
Interactions $g_{\alpha}$ ($\alpha=C,S_{x,y,z},s,t_{x,y,z}$) and
$g_3^{(C)},g_3^{(S)}$ are defined analogously. 
Using these functions as well as the particle-particle 
and particle-hole pair operators defined above, 
and neglecting Umklapp processes, 
we can rewrite the interaction part of the
action in the following way: 
\begin{eqnarray} 
\label{SIgalf} 
S_I &=& -\frac{1}{2} \sum_{\alpha=C,S_{x,y,z}} 
\frac{T}{N} \sum_{k'kq} g_{\alpha}(k',k-q,k'-q,k) 
o^*_{\alpha}(k',q) o_{\alpha}(k,q) \\ 
\nonumber &=& -\frac{1}{2} \sum_{\alpha=s,t_{x,y,z}} 
\frac{T}{N} \sum_{k'kq} g_{\alpha}(k',q-k',q-k,k) 
o^*_{\alpha}(k',q) o_{\alpha}(k,q) \ , 
\end{eqnarray} 
\end{widetext} 
$g_{s,t}$ (resp. $g_{C,S}$) thus describes the interaction between particles 
forming a singlet or triplet pair (resp. particle-hole pair). 
The relation between these couplings is as follows: 
\begin{eqnarray} 
\label{gstvongCS} 
g_s & = & \frac{1}{2} (3 g_S - g_C) \ , \\ 
\nonumber 
g_t & = & \frac{1}{2} (g_S + g_C) \ . 
\end{eqnarray} 
We will make use of these relations in the following, when 
we discuss density fluctuation induced Cooper pairing. Note 
that Umklapp processes do not couple to the Cooper channel, and
therefore affect Cooper pair formation only indirectly \textit{via}
their effect on $g_C$, $g_S$. 

The RG equations are solved numerically, using a fourth order Runge-Kutta 
algorithm \cite{NumericalRecipes} with 
fixed step sizes (taking large steps in the 
beginning of the flow and short steps close to the divergence). 
The graphs showing transition temperatures are
obtained with the help of an adaptive stepsize 
algorithm.\cite{NumericalRecipes,Shampine97} 
As far as the $k\uperp$ dependence of the vertex functions is concerned, 
we discretize the Fermi surface using 32 patches for each sheet. 
Taking advantage of all symmetries of the problem, 
we thus have to calculate 9010 different function values for each 
$\vertex_j$. The use 
of twice the number of patches does not significantly modify our results. 
The integral of the functions $B_{\Cooper,\Peierls}$ must be calculated more 
precisely; we use a fourth order Runge-Kutta algorithm with adaptive 
step size. 

\subsection{\label{local}Results for the case of intrachain interactions} 

To illustrate  our method, 
we shall first consider the by now well known case of 
a quasi-1D electron gas model with purely 
intrachain interactions and no Umklapp scattering. 
This case has been studied in detail by 
Duprat and Bourbonnais \cite{Duprat01} using a Kadanoff-Wilson  
RG scheme  and an approximation where 
only two independent momentum variables for the vertices 
were taken into account. We here use the 1PI scheme and retain 
the full three variable dependence for the vertices. Our results 
confirm those of reference \onlinecite{Duprat01}. 

\begin{figure}
\includegraphics[width=7.0cm]{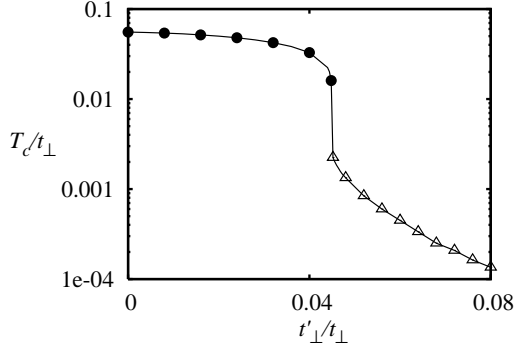}
\caption{\label{localphas}RG phase diagram for the quasi-1D 
electron gas model at incommensurate filling ($g_3=0$) 
with intrachain interactions ($g_j^\perp=0$) and 
nesting deviations, parameterized by $t\uperp'$. 
The circles indicate the transition temperature 
for SDW and the triangles that for SC$d$.  
All figures in this article are obtained using 
the bare intrachain interactions 
\hbox{$\tilde{g}_1=0.32$}, \hbox{$\tilde{g}_2=0.64$} 
and the anisotropy ratio $\Lambda_0/t_\perp=15$.} 
\end{figure}

The RG calculations that follow are performed 
for the values $\tilde{g}_1=0.32$ and $\tilde{g}_2=0.64$, 
which are representative of the couplings likely to be found in practice in 
low dimensional conductors. Thus for not too large 
nesting deviations, the 
renormalization group flow scales to strong coupling, which leads to a  
singular behavior in the susceptibility of a particular 
channel of correlations. This signals an instability of 
the normal state towards 
an  ordered phase. The phase diagram
obtained as a function of $t_\perp'$ is 
shown in Fig.~\ref{localphas}.  For good 
nesting, there is an SDW phase  
with a modulation wave 
vector \hbox{$\mathbf{Q}_0 = (2k_F,\pi)$} that corresponds 
to the best nesting 
vector of the spectrum (\ref{spectrum}). 
The transition temperature obtained for perfect nesting 
($t\uperp'=0$) is $T_c^0 \approx 0.055 t\uperp $. 
If we take \hbox{$t_\perp \approx 200$ K}, we have 
\hbox{$T_c^0 \approx 11$ K}, which falls in the range 
of the experimental $T_c$   
for systems like the Bechgaard salts at low pressure.  
By increasing $t\uperp'$, the transition temperature decreases  
until the  threshold  value 
\hbox{$t_{\perp c}' \approx 0.045 t\uperp \approx 9$ K} is reached, 
where the SDW is suppressed and replaced 
by $d$-wave superconductivity (SC$d$). The maximum 
temperature for the SC$d$ state is found to be 
\hbox{$T_c^0({\rm SC}d)\approx 0.002 t_\perp \approx 0.4$ K}.
These estimates for
$t_{\perp c}'$ and $T_c$ are comparable to the experimental 
results.\cite{Jerome80,Montambaux85,Montambaux91,Duprat01}

\begin{figure}
\includegraphics[width=7.0cm]{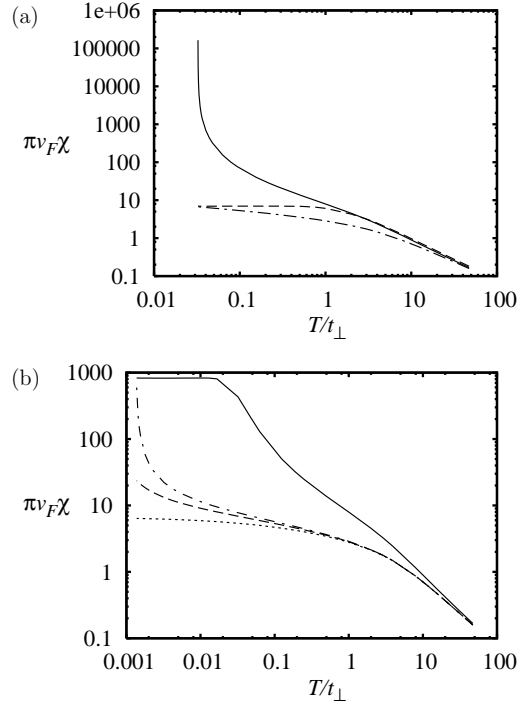}
\caption{\label{chiTs}Temperature dependence 
of the susceptibilities in the normal 
phase ($g_3=0$, $g_j^\perp=0$): \newline 
(a) above the SDW phase ($t\uperp'=0.04t\uperp$). The 
continuous (dashed) line 
corresponds to the SDW response at $q_\perp= \pi \ (0)$; 
the dashed-dotted line to SC$d$ correlations. 
It is interesting to observe that the temperature scale 
$T\sim t_\perp$ below which the $q_\perp=0$ curve separates  
and levels off corresponds to the so-called 
single particle dimensionality crossover.\cite{Bourbon99} 
Our RG scheme thus captures this effect correctly.\newline 
(b) above the SC phase ($t\uperp'=0.048 t\uperp$). 
The continuous line 
corresponds to the SDW response with transverse modulation $\pi$, the 
dashed-dotted line to SC$d_{x^2-y^2}$ ($\cos k\uperp$), the dotted line 
to SC$d_{xy}$ ($r\ \sin k\uperp$) and the dashed line to 
SC$g$ ($r\ \sin 2k\uperp$) correlations.} 
\end{figure} 

The order parameter of the low temperature phase can be 
identified in two different ways, which give equivalent results. 
The first one follows from the identification of the 
most singular behavior in the temperature dependence of 
the various susceptibilities, as shown in Fig.~\ref{chiTs} 
for values of $t_\perp'$ below and above the threshold 
for superconductivity. 

\begin{figure}
\includegraphics[width=7.0cm]{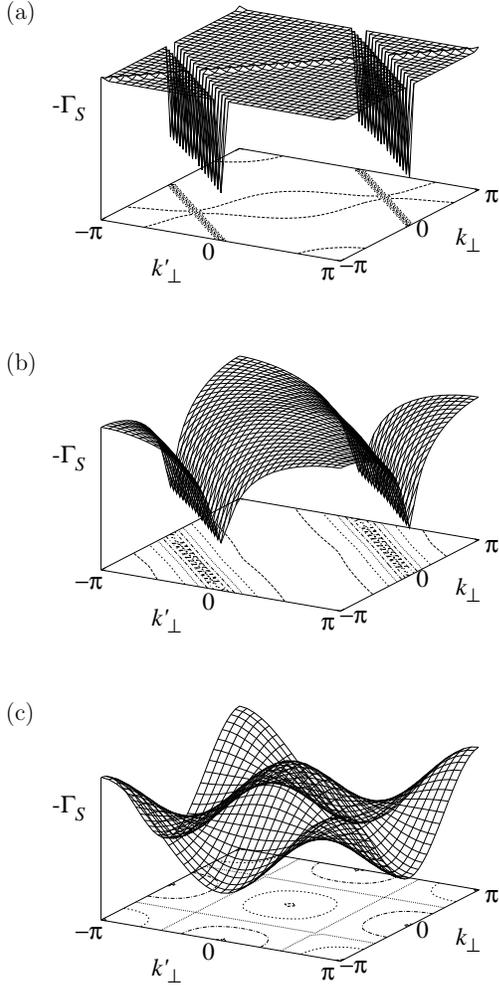}
\caption{\label{gks}$-\tilde{\Gamma}_S(-k\uperp',k\uperp',k\uperp,-k\uperp)$ 
in the SDW regime ($t\uperp'= 0.04 t\uperp$) 
close to the divergence (a), in the SC$d$ regime 
($t\uperp'=0.048t\uperp$) at an intermediate stage of the flow (b) 
and  close to the divergence (c). The vertical scale is arbitrary.} 
\end{figure}

An alternative way to determine the nature 
of the ordered phase is to look 
at the wave vector dependence of the renormalized 
vertex functions close to the divergence. 
In Fig.~\ref{gks}, 
we have plotted the SDW vertex function 
$\vertex_S(-k\uperp',k\uperp',k\uperp,-k\uperp)$ 
in the $k_\perp,k_\perp'$ plane. In the SDW regime, 
only processes involving particle-hole pairs 
at wave vector \hbox{$(-k_\perp')-k_\perp \simeq Q_{\perp 0}$} 
are found to be singular. 
On the other hand, near the superconducting transition temperature, 
we obtain the separable form 
\[ \vertex_S(-k\uperp',k\uperp',k\uperp,-k\uperp) \propto 
\cos k\uperp' \cos k\uperp  \ . \] 
A plot of the
singlet interaction amplitude $\Gamma_s$ gives essentially 
the same picture. From Eq.~(\ref{zstrenorm}), it is 
clear that the most divergent vertex part 
$\orbvertpart$ in such a situation 
is the one proportional to $\cos k\uperp$. 

Recall that the vertex functions $\vertex$ we calculate in our 
RG scheme are equal to the effective interactions $g^{eff}$ of a Wilsonian 
low energy effective theory. Quite early during the flow, 
the vertex associated to the spin density, 
$\vertex_S(k_{j\bot})$, and hence 
$g_S(k_{j\bot})$, develops a peak structure similar to the one close to 
the SDW transition, but less pronounced, cf.\ Fig.~\ref{gks}(b). There 
are thus important spin fluctuations at temperature or energy scales 
above the transition temperature for superconductivity, 
as is confirmed by the behavior of the associated susceptibility,
Fig.~\ref{chiTs}(b). 
The same peak structure appears in the effective interaction between 
electrons forming a 
singlet (or triplet) pair, $g_{s(t)}(k_{j\bot})$. 
We can decompose the peak of 
$g_s(-k\uperp',k\uperp',k\uperp,-k\uperp)$ at 
intermediate cutoff $\Lambda$ in terms of 
the variables $(k\uperp+k\uperp')$ and $(k\uperp-k\uperp')$. Neglecting the weak $(k\uperp-k\uperp')$ dependence, the result is, schematically, 
\begin{eqnarray}
\label{spinfluktpeak} 
&& g_s (-k\uperp',k\uperp',k\uperp,-k\uperp) \\ 
\nonumber  
&=& - \frac{1}{N\uperp} \sum_{n=-\frac{N\uperp}{2}+1}^{\frac{N\uperp}{2}} 
a_n \, e^{i(k\uperp'+k\uperp-\pi)n} \\ 
\nonumber  
&=& - \frac{1}{N\uperp} \sum_{n=-\frac{N\uperp}{2}+1}^{\frac{N\uperp}{2}} 
a_n \, (-1)^n \\ 
\nonumber && 
\times \bigl( \cos nk\uperp' \cos nk\uperp 
- \sin nk\uperp' \sin nk\uperp \bigr) \ . 
\end{eqnarray} 
$\betrag{n}$ corresponds to the distance 
between the chains where the two 
interacting electrons are located. 
In the SDW regime, Fig.~\ref{gks}(a), $g_s$ has the form of a $\delta$
peak, so that all $a_n>0$ will be equal. 
In the superconducting regime, Fig.~\ref{gks}(b), the peak is slightly 
enlarged, so that $a_n>0$ will be some decreasing function of 
$\betrag{n}$. We thus see that, due to the spin fluctuations, 
the effective interaction between particles forming 
a singlet pair contains attractive contributions 
at all chain distances. (Note that, according to our definitions 
Eqs.~(\ref{CSstvertexdef}) and (\ref{SIgalf}), 
an ``attractive'' interaction corresponds to 
\textit{positive} $g_{s,t}$.) 
The most important one is the nearest-neighbor chain one, 
$\cos k\uperp$ (SC$d$), 
followed by $\sin 2k\uperp$ (SC$g$), $\cos 3k\uperp$ etc., as it can 
also be seen from the calculation of the related pairing
susceptibilities, cf.\ Fig.~\ref{chiTs}(b). 
Note that, according to the relations~(\ref{gstvongCS}), 
$g_S$ also gives an attractive
contribution to the triplet channel. However, 
all the three components of a  -- spin-one boson -- SDW fluctuation 
contribute to the superconducting coupling in the singlet 
channel, whereas only one contributes to the triplet channel. 
Antiferromagnetic fluctuations thus favor singlet 
pairing as compared to triplet pairing. 
On the other hand, according to
Eq.~(\ref{gstvongCS}), strong charge fluctuations should in 
principle be able to change this tendency.

\section{\label{innter}Interchain interactions and 
triplet superconductivity: Incommensurate case} 

In this section, we examine the role of interchain interactions in 
the phase diagram of the extended quasi-1D electron gas 
model. We will first 
consider the influence of interchain backward scattering $\gnn_1$ and 
forward scattering $\gnn_2$ separately, before studying their combined
effect. The intrachain interactions are kept fixed to their values used 
in the previous section. 
Concerning the nesting quality,
the overall picture remains the same. For weak deviations from perfect
nesting, we find density-wave instabilities, superconductivity for
more important deviations, and -- discarding possible Kohn-Luttinger effects
at low temperatures beyond the numerical accuracy of our RG calculations -- 
a metallic phase when the nesting is deteriorated even more. 

\begin{figure}
\includegraphics[width=7.0cm]{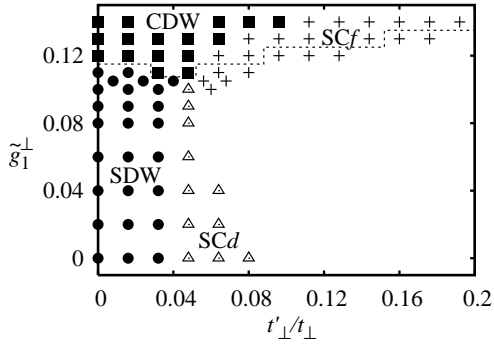}
\caption{\label{g1nnphas}Low temperature phases versus 
$\gnn_1$, keeping $\gnn_2=0$ and $\tilde{g}_3=0$. 
Circles indicate a SDW phase, squares a CDW phase, triangles SC$d$ 
($\cos k\uperp$)
and crosses SC$f$ ($r\ \cos k\uperp$). In the region below 
the dotted line, spin fluctuations dominate 
over charge fluctuations in the normal phase.} 
\end{figure} 

\begin{figure}
\includegraphics[width=7.0cm]{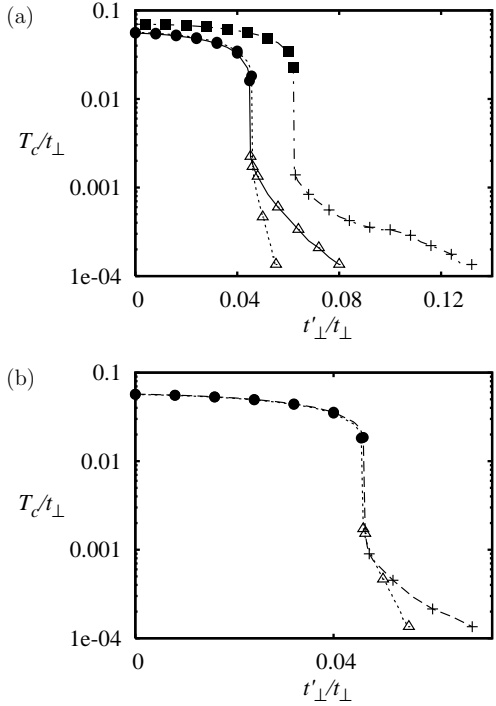}
\caption{\label{Tcbann}Transition temperatures 
for different values of $\tildegnn_1$, 
when $\tildegnn_2=0$ and $\tilde{g}_3=0$: \newline 
(a) $\tildegnn_1=0$ (phase sequence
SDW$\rightarrow$SC$d$, continuous line), 
$0.090$ (SDW$\rightarrow$SC$d$, dotted line) and
$0.120$ (CDW$\rightarrow$SC$f$, dashed-dotted line); \newline 
(b) $\tildegnn_1=0.090$ (SDW$\rightarrow$SC$d$, dotted line) and 
$0.105$ (SDW($\rightarrow$SC$d$)$\rightarrow$SC$f$, dashed line). 
Note that in the latter case, the SC$d$ phase is extremely narrow.} 
\end{figure} 

\begin{figure}
\includegraphics[width=7.0cm]{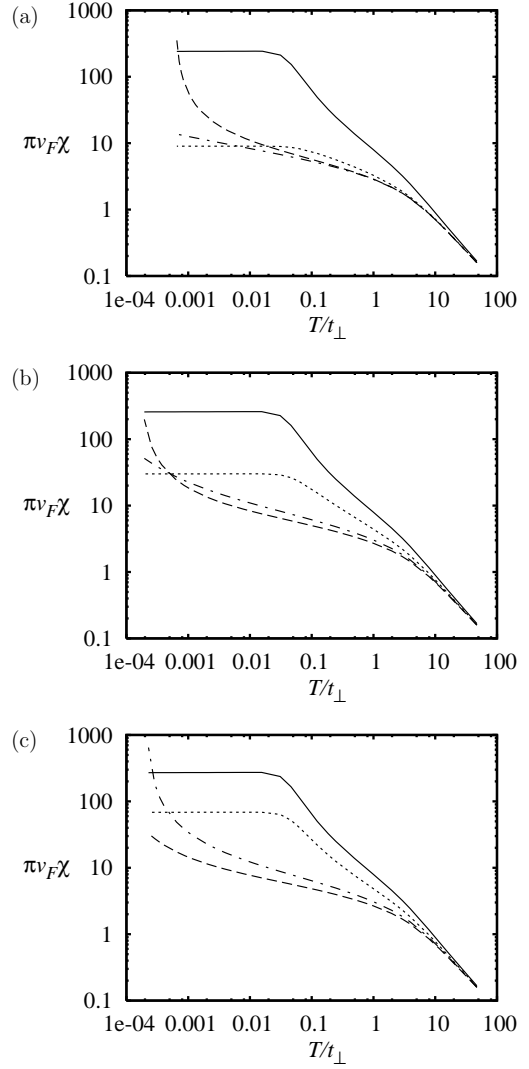}
\caption{\label{chiTvgl}Temperature dependence 
of the susceptibilities in the normal phase 
above the SC$d$ phase ((a) $\tilde{g}_1^\perp=0$ and 
(b) $\tilde{g}_1^\perp=0.08$) and above the SC$f$ phase 
((c) $\tilde{g}_1^\perp=0.10$), keeping $\tilde{g}_2^\perp=0$, 
$\tilde{g}_3=0$ and $t_\perp'=0.056 t_\perp$. 
The continuous line corresponds to SDW, the dotted line to CDW, 
the dashed line to SC$d$ and the dashed-dotted line to SC$f$ 
correlations.} 
\end{figure}

By switching  on $\gnn_1$ gradually on the positive scale, one 
observes that at small values of $\gnn_1$, the SDW phase remains 
essentially unaffected, whereas the transition temperature for $d$-wave 
superconductivity is considerably reduced by the presence of 
a finite $\gnn_1$ (Fig.~\ref{Tcbann}), 
and the region where superconductivity is stable
shrinks (Fig.~\ref{g1nnphas}). 
For higher $\gnn_1$ ($\gnn_1$ of the order of 
$\frac{1}{3}$ of the intrachain backscattering $g_1$), $d$-wave  
superconductivity turns out to be no longer stable and an $f$-wave 
triplet superconducting phase appears. 
In this parameter range, $(2k_F,\pi)$ charge 
fluctuations are strongly enhanced by  
interchain backward scattering 
(Fig.~\ref{chiTvgl}). However, 
spin fluctuations remain important, 
and in a sizeable region of the phase 
diagram, triplet superconductivity is preceded 
in temperature by dominant spin fluctuations in the normal 
state (Figs.~\ref{g1nnphas} and \ref{chiTvgl}). 
For the values of intrachain 
interactions used in this section, and $\tilde{g}_1^\perp=0.105$, 
one finds a maximum 
$T_c^0({\rm SC}f) \approx 0.001 t_\perp \sim 0.2$~K, 
that is of the same order of
magnitude as that of the $d$-wave case. $T_c({\rm SC}f)$ increases 
with the amplitude of $\gnn_1$, 
and the superconducting phase widens (Figs.~\ref{g1nnphas} and
\ref{Tcbann}). Once triplet superconductivity occurs, 
$\gnn_1$ starts to affect the density
wave phase: The SDW state 
is suppressed and replaced by a CDW. 
The values of $\tilde g_1^\perp$ for which SC$f$ and CDW phases first
appear depend 
on the values of intrachain interactions and increase with 
the value of the ratio  $g_1/g_2$. 

\begin{widetext} 

The origin of the $f$-wave SC and CDW phases
can be understood by considering the contribution of the
$g^\perp_j$'s to the (bare) scattering
amplitudes in the singlet and triplet particle-particle 
channels, as well as in the charge and spin channels: 
\begin{eqnarray} 
\label{gnnCSvongnn12} 
\gnn_C(k\uperp'+q\uperp,k\uperp-q\uperp,k\uperp',k\uperp) & = & 
-4\gnn_1\cos q\uperp + 2\gnn_2 \cos (q\uperp+k\uperp'-k\uperp) \ , \\ 
\nonumber
\gnn_S(k\uperp'+q\uperp,k\uperp-q\uperp,k\uperp',k\uperp) & = & 
2\gnn_2 \cos (q\uperp+k\uperp'-k\uperp) \ , 
\end{eqnarray} 
\begin{eqnarray} 
\label{gnnstvongnn12} 
\gnn_s(-k\uperp',k\uperp',k\uperp,-k\uperp) & =& 
2 (-\gnn_1-\gnn_2) \cos k\uperp' \cos k\uperp 
+ 2 (\gnn_1-\gnn_2) \sin k\uperp' \sin k\uperp \ , \\ 
\nonumber 
\gnn_t(-k\uperp',k\uperp',k\uperp,-k\uperp) & = & 
2 (\gnn_1-\gnn_2) \cos k\uperp' \cos k\uperp 
+ 2 (-\gnn_1-\gnn_2) \sin k\uperp' \sin k\uperp \ . 
\end{eqnarray} 
From these equations, it can be easily seen
that the interchain repulsion $\gnn_1$ 
contributes positively to $g_C^\perp$ at $q\uperp=\pi$, and therefore
induces CDW correlations with a phase difference of $\pi$ between
neighboring chains. In the Cooper channel, $g_1^\perp$ 
favors triplet $f$-wave and singlet $d_{xy}$-wave
pairing, whereas its contribution to singlet $d_{x^2-y^2}$-wave and
triplet $p_y$-wave pairing is negative. As for $g_2^\perp$, it tends to
suppress both singlet and triplet pairings on nearest-neighbor chains. 

In addition to this `direct' contribution there is also an indirect effect due
to the exchange of density fluctuations. Upon renormalization, 
$(2k_F,\pi)$ CDW correlations are enhanced beyond the level 
expected from a mean-field treatment of  
$\gnn_1$.\cite{Menyhard77} These CDW fluctuations enhance
triplet $f$-wave pairing but suppress singlet pairing, whereas 
the SDW fluctuations are well known to favor singlet
pairing (see Eqs.~(\ref{gstvongCS})). Eq.~(\ref{spinfluktpeak}) shows that
the latter reinforce $d_{x^2-y^2}$- but suppress $d_{xy}$-wave pairing. 
Regardless of the values of the interaction constants
$g_1$, $g_2$ and $g_2^\perp$, the CDW and triplet $f$-wave phases always 
appear almost simultaneously when $g_1^\perp$ increases. This suggests that
CDW fluctuations (rather than the direct effect of $g_1^\perp$ in the
Cooper channel [Eqs.~(\ref{gnnstvongnn12})]) provide the dominant driving
force leading to $f$-wave superconductivity. 
\end{widetext} 

\begin{figure}
\includegraphics[width=7.0cm]{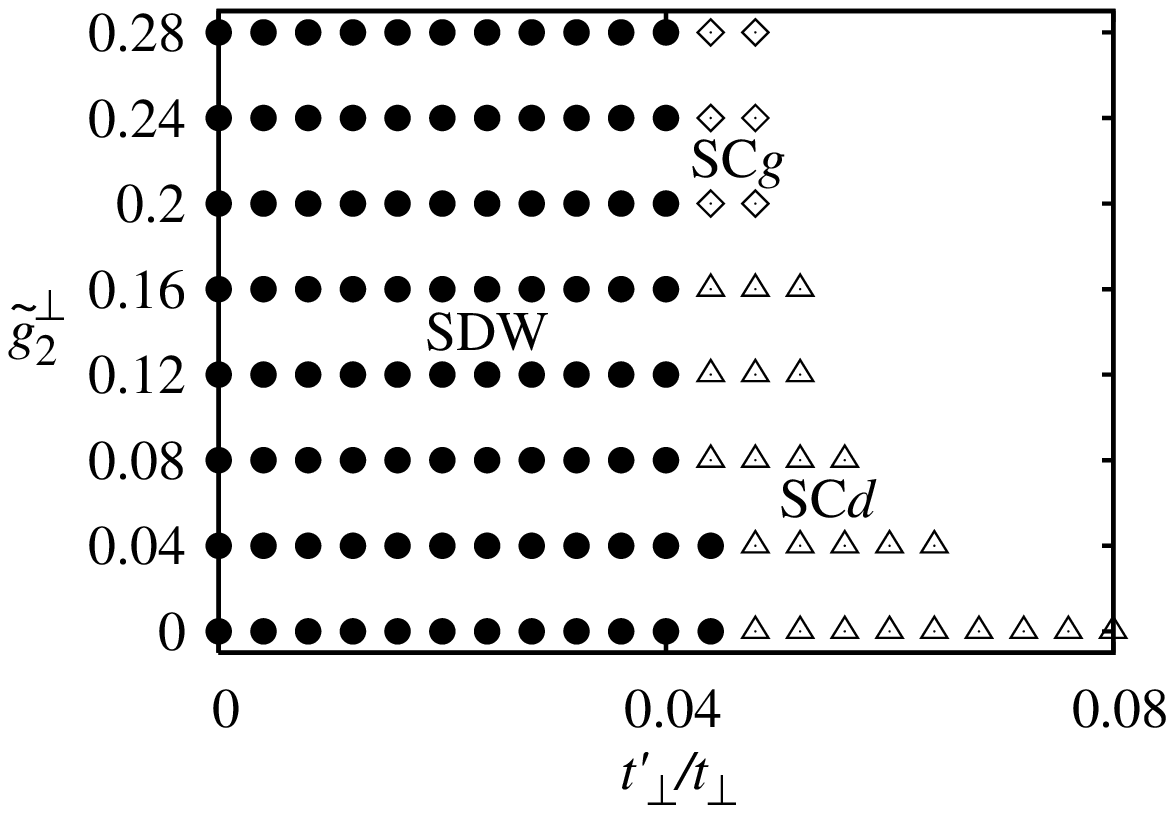}
\caption{\label{g2nnphas}Low temperature phases for 
$\gnn_2$ varying, keeping $\gnn_1=0$ and $\tilde{g}_3=0$. 
Circles indicate a SDW phase, triangles SC$d$ 
and diamonds SC$g$ ($r\ \sin 2k\uperp$).}  
\end{figure} 

\begin{figure}
\includegraphics[width=7.0cm]{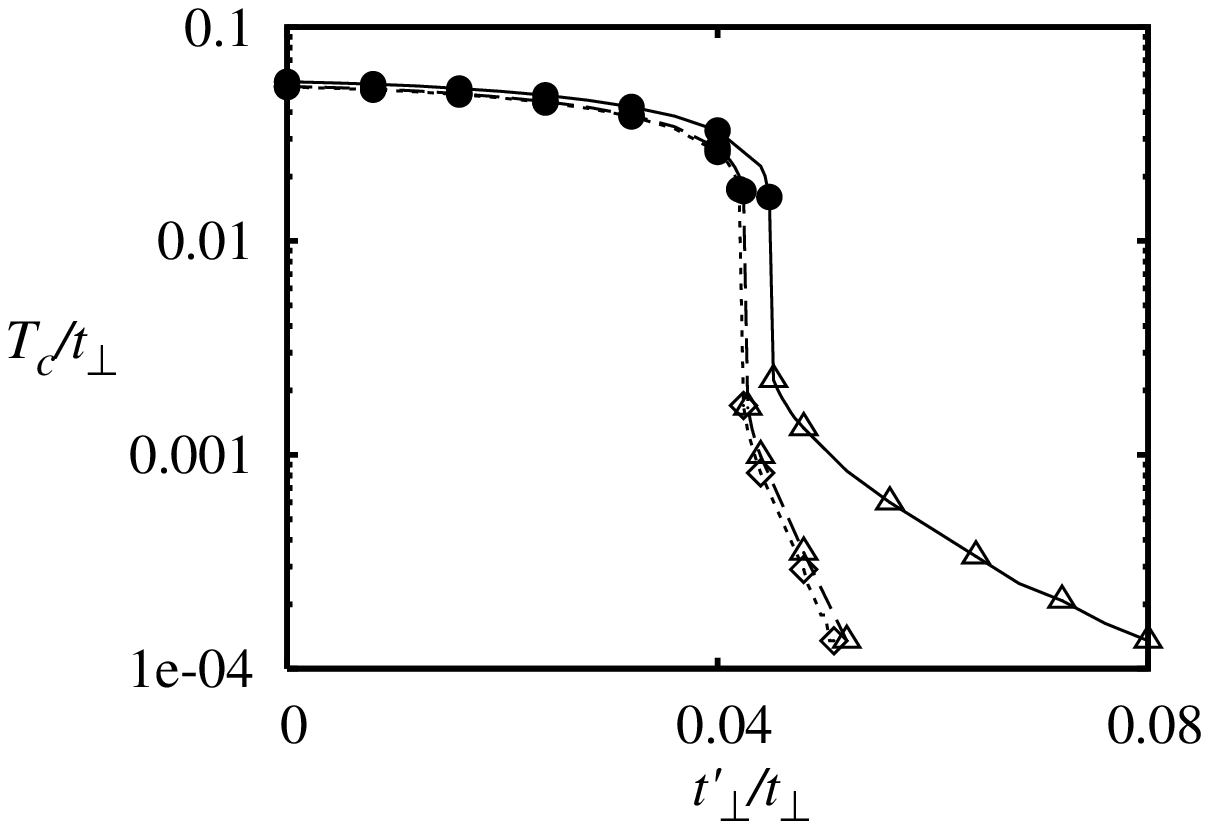}
\caption{\label{Tcfonn}Transition temperatures 
for different values of $\tildegnn_2$, if $\tildegnn_1=0$ and
$\tilde{g}_3=0$: 
$\tildegnn_2=0$ (SDW$\rightarrow$SC$d$, continuous line), 
$0.16$ (SDW$\rightarrow$SC$d$, dashed line) 
and $0.20$ (SDW$\rightarrow$SC$g$, dotted line).} 
\end{figure}

We now consider the effect of interchain \textit{forward }scattering
$g_2^\perp$ alone, setting $\gnn_1=0$. 
From Eqs.~(\ref{gnnstvongnn12}), it can
easily be seen that $\gnn_2$ contributes negatively to 
both $g_s^\perp$ and $g_t^\perp$ and then favors the 
suppression of the nearest-neighbor-chain Cooper pairing 
induced by spin fluctuations. Our results (Figs.~\ref{g2nnphas} and 
\ref{Tcfonn}) show that this is indeed the case. We have
seen, however, in section~\ref{local}, that SDW fluctuations can
generate a smaller 
yet present attractive interaction between electrons on
next-nearest-neighbor chains, which is not affected by
$\gnn_2$.  It follows that when $d$-wave superconductivity 
is sufficiently weakened by
$\gnn_2$, it is  replaced by $g$-wave singlet 
pairing ($\Delta_r(k_\perp) \propto r\, \sin 2k\uperp$). This is shown in 
Fig.~\ref{g2nnphas}. 

It is worth noting that 
the instability of the normal state with respect to superconductivity with
high angular momentum 
pairing can be seen as a result of the Kohn-Luttinger 
effect originally predicted for isotropic metals.\cite{Kohn65} 
At variance with more isotropic systems, however, the RG results
show that for a quasi-1D metal, the transition 
temperature of high angular momentum superconducting phases, 
like SC$g$, remains experimentally accessible. We find
$T_{c,SCg}^{max} \sim 0.002 t\uperp \sim 0.4\ {\rm K}$. Finally, the
SDW phase remains nearly unaffected by $\gnn_2$. 
This can be understood from 
Eqs.~(\ref{gnnCSvongnn12}), which indicate 
that its contribution averages out over the Fermi surface 
in the particle-hole channel of the RG equations. 

\begin{figure}[h]
\includegraphics[width=7.0cm]{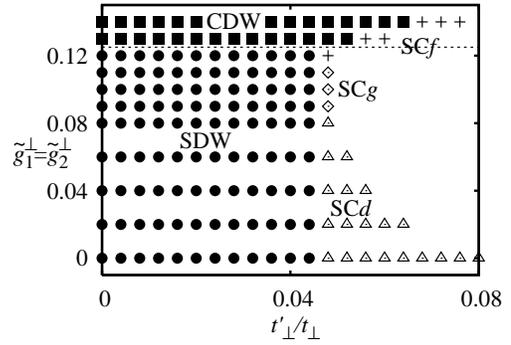}
\caption{\label{diagphas}Low temperature phases for 
$\gnn_1=\gnn_2$, $g_3=0$. 
Circles indicate a SDW phase, black squares a CDW phase, triangles SC$d$ 
($\cos k\uperp$), white diamonds SC$g$ ($r\ \sin 2k\uperp$) 
and crosses SC$f$ ($r\ \cos k\uperp$). Below the dotted line, 
spin fluctuations dominate over charge fluctuations in the normal phase.} 
\end{figure}

We now consider the combined effect of $\gnn_1$ and
$\gnn_2$. Fig.~\ref{diagphas} shows the results for $\gnn_1=\gnn_2$, 
where all
aforementioned phases appear. In the presence of both $\gnn_1$ and
$\gnn_2$, $d$-wave pairing is  suppressed even faster than
by $\gnn_1$ or $\gnn_2$ alone. The appearance of $f$-wave pairing is
retarded by $\gnn_2$ which, as previously mentioned, 
is detrimental to nearest neighbor chain pairing
(Eq.~(\ref{gnnstvongnn12})). However, 
for our choice of intrachain interactions, no triplet phase with 
pairing on next-nearest-neighbor chains 
is found for $\gnn_1=\gnn_2$, since $\gnn_1$ favors nearest-neighbor 
triplet pairing at the outset.
As for CDW's, they are found to occur
at slightly higher values of $\gnn_1$ in the presence of a finite
$\gnn_2$. As mentioned before,
$\gnn_2$ has no important effect in the particle-hole channel
alone. The slight suppression of the CDW due 
to $\gnn_2$ must therefore come from the 1D 
regime, where the correlation channels are coupled. 
This can be checked from the RG
equations for $t\uperp=0$ given in appendix~\ref{glims}.

The phase diagram (Fig.~\ref{diagphas}) depends quantitatively and 
qualitatively on the
bare intrachain interactions. When the ratio $g_1/g_2$ 
increases, higher values of the interchain interactions are necessary
for obtaining CDW and SC$f$ phases because 
$\Gamma_C$ is more negative at the outset. 
For $g_1=g_2$, we even find 
a triplet SC$f$ order parameter of the form 
$\Delta_r(k_\perp)\propto \sin 2k_\perp$, corresponding to
  second-nearest-neighbor chain pairing, instead of 
$\Delta_r(k_\perp)\propto r\cos k_\perp$. 
This is coherent with the fact that, in the 1D 
regime, the renormalization of $\Gamma_1^\perp$ is proportional 
to $\Gamma_1^\perp \Gamma_C$, so that a more negative $\Gamma_C$ 
reduces $\Gamma_1^\perp$ and hence nearest-neighbor chain 
triplet pairing (appendix \ref{glims}). 

\section{\label{umklapp}Effect of Umklapp processes} 

Conductors like the Bechgaard and Fabre salts
are slightly dimerized in the direction
of the organic chains. It follows that at low energy or temperature,
the hole band can be considered as effectively half filled rather
than quarter-filled and this gives rise 
to Umklapp scattering processes
with amplitudes $g_3$ and $g_3^\perp$.
To leading order, the
bare amplitude of $g_3\approx g_1 \Delta_D/E_F$ 
is proportional to the
dimerization gap $ \Delta_D$,\cite{Barisic81,Emery82,Penc94} which
yields a $g_3$ that is rather weak
as compared to $g_{1,2}$. We assume a similar
ratio between the interchain
couplings $g_3^\perp$ and $g_{1,2}^\perp$.\cite{note1} 

\begin{figure}
\includegraphics[width=7.0cm]{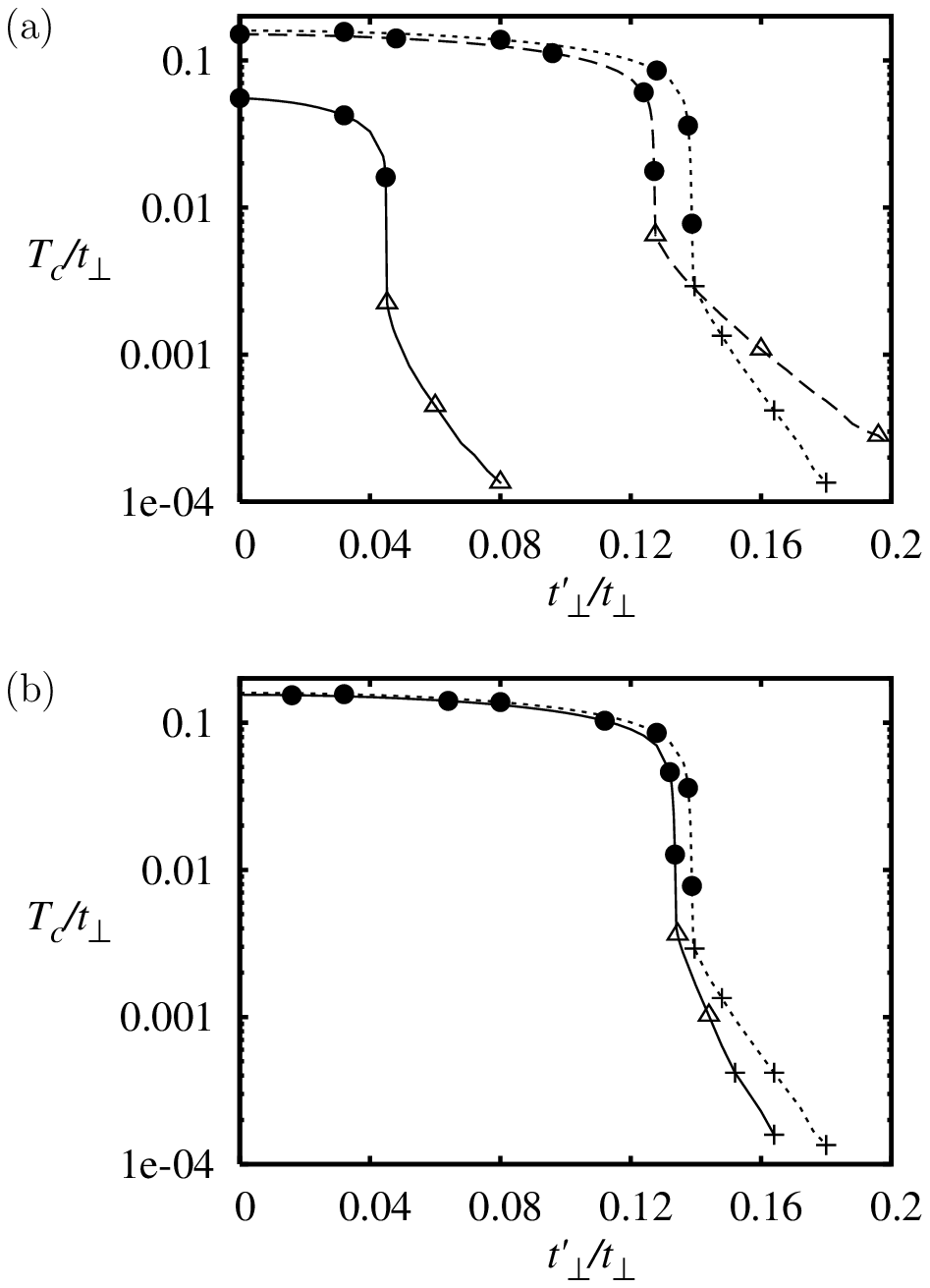}
\caption{\label{umTc}Transition temperatures 
when Umklapp processes are taken into account.\newline 
(a) Without interchain Umklapp scattering: 
$\tilde{g}_3=0$, $\tilde{g}_j^\perp=0$ (SDW$\rightarrow$SC$d$,
continuous line), \newline 
$\tilde{g}_3=0.02$, $\tilde{g}_j^\perp=0$ (SDW$\rightarrow$SC$d$,
dashed line), \newline 
$\tilde{g}_3=0.02$, $\tilde{g}_1^\perp=\tilde{g}_2^\perp=0.1$, 
$\tilde{g}_3^\perp=0$, (SDW$\rightarrow$SC$f$, dotted line). \newline 
(b) Effect of interchain Umklapp processes: 
$\tilde{g}_3=0.02$, $\tilde{g}_1^\perp=\tilde{g}_2^\perp=0.1$, 
$\tilde{g}_3^\perp=0$ (SDW$\rightarrow$SC$f$, dotted line) and
\newline 
$\tilde{g}_3^\perp=\frac{\tilde{g}_1}{\tilde{g}_3}\tilde{g}_1^\perp$ 
(SDW$\rightarrow$SC$d$$\rightarrow$SC$f$, continuous line).} 
\end{figure} 

\begin{figure}[h]
\includegraphics[width=7.0cm]{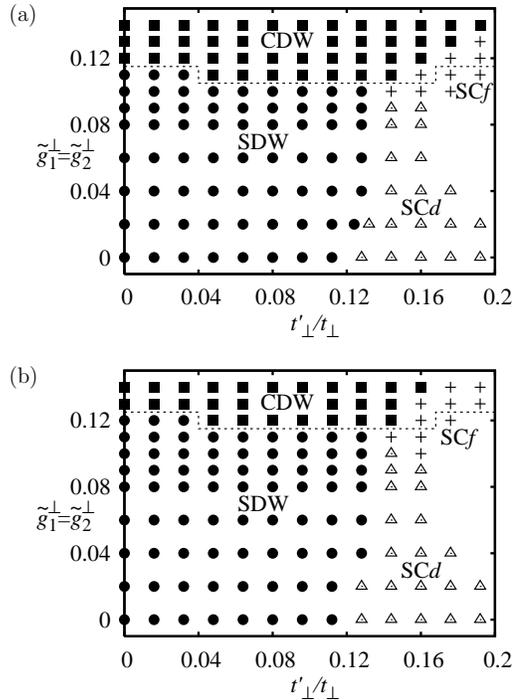}
\caption{\label{umklphas}Low temperature phases 
in the presence of Umklapp processes,
with $\gnn_3=0$ (a) and $\gnn_3/g_3=\gnn_1/g_1$ (b), 
taking $\tilde g_2^\perp=\tilde g_1^\perp$. 
Circles indicate a site-SDW phase, squares a bond-CDW phase, 
triangles SC$d$ ($\cos k\uperp$) and 
crosses SC$f$ ($r\ \cos k\uperp$). Below the dotted lines, spin
fluctuations dominate over charge fluctuations in the normal phase.} 
\end{figure} 

\begin{figure}
\includegraphics[width=7.0cm]{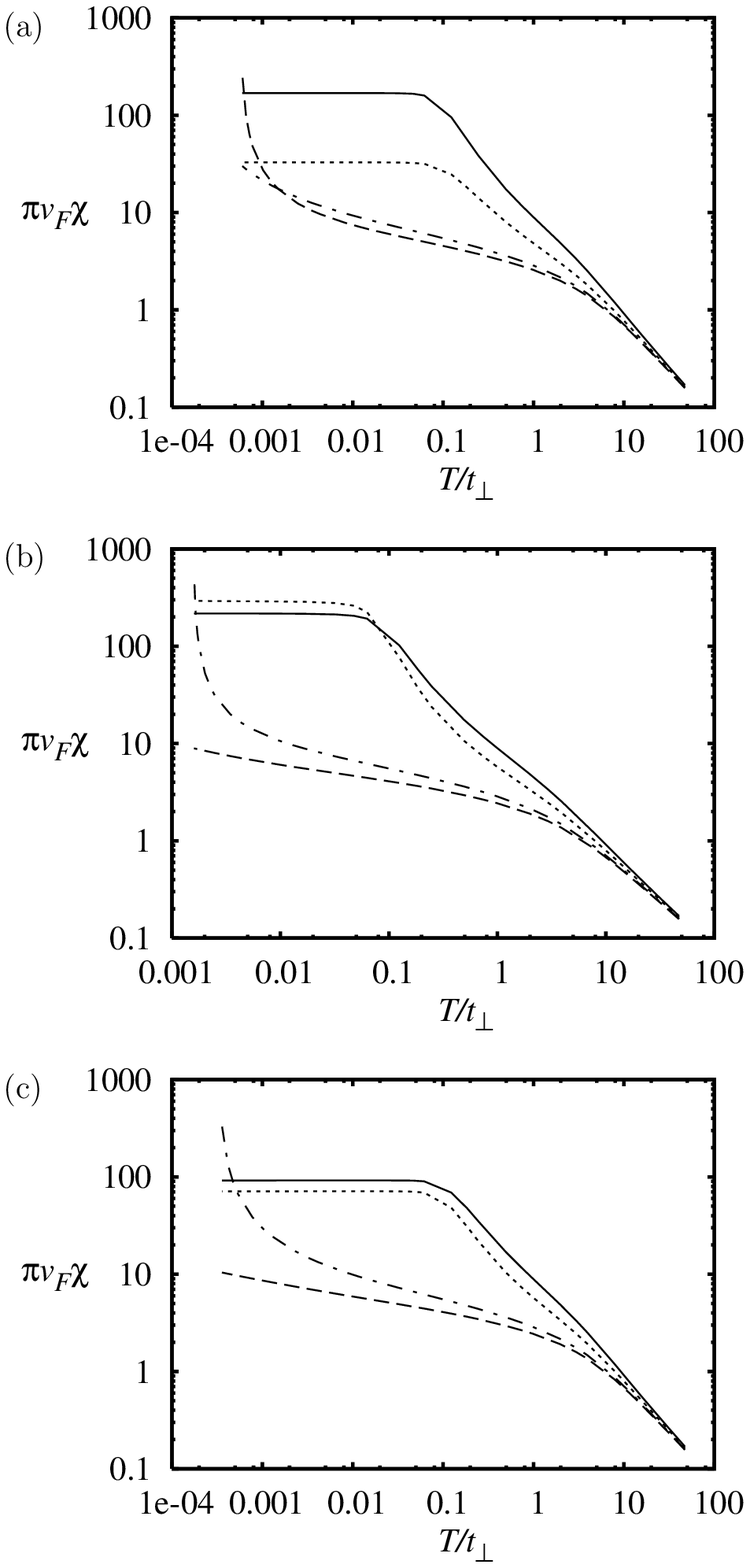}
\caption{\label{umchi3in1}Temperature dependence 
of the susceptibilities in the normal phase 
above the SC$d$ phase ((a) $t_\perp'=0.152 t_\perp$, 
$\tilde{g}_1^\perp=0.08$) and above the SC$f$ phase 
((b) $t_\perp'=0.152 t_\perp$ and (c) $t_\perp'=0.176 t_\perp$, 
both for $\tilde{g}_1^\perp=0.12$), taking $\tilde{g}_3=0.02$ and 
$\tilde{g}_2^\perp=\tilde{g}_1^\perp=
\frac{\tilde{g}_1}{\tilde{g}_3}\tilde{g}_3^\perp$. 
The continuous line corresponds to SDW, the dotted line to CDW, 
the dashed line to SC$d$ and the dashed-dotted line to SC$f$ 
correlations.} 
\end{figure} 

Intrachain Umklapp processes 
enhance the formation of `site' centered 
SDW correlations and `bond' centered CDW correlations  
and weaken bond SDW and site CDW fluctuations, as can be seen from
Eqs.~(\ref{zDWrenorm}).  
Thus, contrary to the incommensurate case, site and bond 
density-wave orders no longer join together and must be 
considered separately with their distinct strengths at 
half filling. In addition, intrachain Umklapp scattering reinforces
the effective interactions $g_C$ and $g_S$ at 
momentum transfers $(2k_F,\pi)$, 
thus increasing spin and charge density fluctuations of both site
and bond type. These effects are known in the 1D case 
(see \textit{e.~g.\ }Ref.~\onlinecite{Bourbon95}
and also Appendix~\ref{glims}), as well as from calculations restricted to
the particle-hole channels (Ref.~\onlinecite{Dupuis98b} and
Appendix~\ref{glims}). 
Even though Umklapp processes do not enter the particle-particle 
channel directly, they do enhance Cooper pairing, 
because they amplify the peak 
at $(2k_F,\pi)$ in the effective interactions $g_{S,C}$ and 
thus the attraction between electrons on neighboring chains
[see Eq.~(\ref{spinfluktpeak})]. 
Since density-wave correlations are
enhanced with respect to the incommensurate case, transition
temperatures are also higher.\cite{Bourbon04} This is true for the
formation of density-wave states as well as for superconductivity 
(Fig.~\ref{umTc}(a)). 
The ratio $T_c^0/T_{c,SC}^{max}$ remains essentially
unaffected by $g_3$. 
The critical $t_{\perp c}'$ needed to destroy the density-wave phase is also
higher in the presence of Umklapp scattering. 

Let us now turn to the effect of interchain interactions in the
presence of Umklapp processes. We will first consider the 
effect of a finite intrachain $g_3$, for 
$\gnn_1=\gnn_2>0$ and $\gnn_3=0$.
A comparison of
Figs.~\ref{diagphas} and \ref{umklphas}(a) shows that the regions of
$d$-wave and $f$-wave superconductivity are now enlarged, and there is
no more SC$g$ phase corresponding to 
next-nearest-neighbor chain singlet pairing. This is a
consequence of the fact that density-wave correlations are reinforced by
Umklapp scattering. A stronger $\gnn_2$ would then be needed to
destroy nearest-neighbor chain pairing, but owing to the 
condition $\gnn_1=\gnn_2$
the SC$f$ phase is also reinforced
and in turn stabilized once the SC$d$ phase is suppressed. 

When we finally add interchain Umklapp scattering $\gnn_3$, the
picture does not change significantly (Fig.~\ref{umklphas}(b)). The
only effect is that the occurrence of SC$f$ 
and CDW phases takes place at slightly higher values of the
interchain interactions in the phase diagram, 
whereas transition temperatures are scarcely 
lower (Fig.~\ref{umTc}(b)). In the 1D
regime ($\Lambda\gg t_\perp$), $\gnn_3$ enhances $g_C$
(appendix~\ref{glims}), but this is only a second order effect. On the
other hand, the bare contribution of $\gnn_3$,
Eq.~(\ref{gstarting}), reduces the peak in the effective interaction 
$g_3(k_{1\bot}',k_{2\bot}',k_{2\bot},k_{1\bot})$ at
$k_{1\bot}'-k_{1\bot}=\pi$, which is connected to the charge density 
fluctuations (appendix~\ref{glims}). This influence is linear 
in $\gnn_3$ and therefore stronger
than the 1D effect, thus explaining the overall weakening
of CDW  and, in turn, of SC$f$
correlations. There is no direct effect of $\gnn_3$ on SDW 
correlations. However, the RG equations
(Appendix~\ref{glims}) show that $\gnn_3$ weakens the renormalization
of $g_3$ in the 1D regime and in turn the increase of transition
temperatures in comparison to the incommensurate situation. 

In Fig.~\ref{umchi3in1}, we show the typical behavior of the most
important susceptibilities as a function of temperature in the normal
phase above the superconducting phases. Fig.~\ref{umchi3in1}(a) shows 
that for sufficiently important interchain interactions, triplet
correlations are already strongly enhanced, although the ordered phase
still corresponds to spin singlet pairing. SDW
correlations are always important. They are the dominant 
fluctuations in the normal
phase in the major part of the parameter range we have explored, 
sometimes even above the triplet SC phase 
(cf.\ Figs.~\ref{umchi3in1}(c) and \ref{umklphas}).

\section{\label{nnandexp} Discussion and Conclusion} 

In this work, we have determined the possible electronic phases 
of the extended quasi-1D electron gas model  that includes  both  
intrachain and interchain repulsive interactions, interchain 
electron hopping and  the influence of nesting deviations.  
Our results reveal that in correlated quasi-1D metals both for 
zero and non zero Umklapp scattering,  interchain interactions 
can act  as  a key factor in expanding the range of  possibilities 
of ordered states compared to the case where only repulsive 
intrachain interactions are present.\cite{Duprat01,Bourbon04,Fuseya05} 
At large momentum transfer,  the interchain electron-electron coupling 
acts as a short-range interaction that is responsible for the
enhancement of CDW correlations, consistently with what was found 
long ago in the absence of interchain hopping.\cite{Gorkov74,Lee77} 
For finite $t_\perp$, however,  interchain interaction  leaves the 
amplitude of SDW correlations essentially unaffected and a 
relatively small critical value of $g_1^\perp$ coupling is  
then needed to make the CDW ordered state possible, and this, 
on equal footing with the SDW  phase, which is known to dominate  
the phase diagram   when  only repulsive intrachain interactions 
are present.  When nesting deviations  are cranked up beyond some 
threshold value $t_{\perp c}'$, density-wave order is suppressed 
and  interchain interactions turn out to affect correlations of the 
Cooper channel too. Thus an important conclusion that  emerges from 
this work is the gradual suppression of interchain $d$-wave pairing  
when  repulsive  short-range  interchain interaction increases. 
As a result of the growth of CDW correlations  in the normal 
state, it  ultimately yields the stabilization of a 
triplet SC$f$ superconducting  phase corresponding to an order 
parameter $\Delta_r(k_\perp) = r\Delta \cos k_\perp $ having nodes 
on the warped Fermi surface that are at the same locus as for 
the SC$d$ case. The normal phase  is still dominated
by strong SDW fluctuations over a sizable region of the phase diagram in this
sector. 

It is interesting to consider how far the RG results of this 
work are applicable to quasi-1D organic conductors. In the 
case of the Bechgaard (TMTSF)$_2$X salts with
centrosymmetric anions X for example, the  observation of a 
SDW-CDW  coexistence below the  critical pressure for 
superconductivity\cite{Ng84,Pouget96,Kagoshima99} indicates 
that interchain electron repulsion is a relevant interaction 
in these materials besides intrachain interactions and 
interchain hopping. Accordingly, the RG calculations show 
that a relatively  small and realistic  amplitude of repulsive 
interchain interaction is sufficient to bring the stability 
of CDW order close to SDW, indicating that this part of 
interaction  would  indeed  play  an important role in 
the emergence of density-wave order in these materials. 
The suppression  of the SDW state (and presumably of CDW 
as well)   followed by the emergence of superconductivity  
is well known to constitute the closing sequence of 
transitions that characterizes virtually all members of 
the Bechgaard and Fabre salts series as one moves along the 
pressure scale. Since pressure introduces alterations of the 
electron spectrum, it  prompts   deviations  from perfect 
nesting. In our  model,  these simulate  the main effect 
of pressure, which together with a reasonable set of 
parameters, yield  a `pressure' profile of the critical 
temperature that agrees quite well with the  
characteristic variation seen in experiments.\cite{Jerome82} 
As  to the nature of the superconductivity in these materials, 
our results  based on a purely electronic model,  indicate that 
given the observation of the close proximity between SDW and CDW 
ordered states, not only SC$d$ but also triplet SC$f$ order 
parameter become   serious candidates for the description 
of the superconducting phase in these compounds (Figs.~\ref{g1nnphas},\ref{diagphas}, and \ref{umklphas}). 
However, as pressure also affects the normalized amplitudes 
$\tilde{g}^{(\perp)}_i$ through the band width and the 
dimerization gap (Umklapp), and  $g_i^\perp$ through the 
interchain distance,\cite{Saub76,Barisic81} the actual 
trajectory in the phase diagram under pressure cannot be 
determined with great precision. It follows that besides 
the possibilities SDW$\to$ SC$d$ or SDW $\to$ SC$f$, 
sequence  of transitions such as  SDW $\to$ SC$d$ $\to$ SC$f$, 
where one can pass from singlet to triplet SC
order under  pressure, cannot be excluded. It is worth 
remarking that in this sector of the phase diagram, the 
addition of a small magnetic field -- as actually used 
in many experiments\cite{Lee02,Oh04} --  or accounting 
for the small but yet finite  spin anisotropy  would 
tend to tip the balance in favor of a triplet order 
parameter.\cite{Shimahara00,Fuseya05} 

Experimental features of the normal phase also argue 
in favor of this region  of the phase diagram for the 
Bechgaard salts. This is the case of the puzzling growth 
of  CDW correlations seen in optical conductivity in  the 
low temperature part of the metallic phase above the 
superconducting transition.\cite{Cao96} CDW correlations 
are found to be  significantly  enhanced in a temperature 
region where NMR experiments reveal the existence of strong 
SDW correlations.\cite{Bourbon84,Wzietek93} This feature 
cannot be captured for realistic intrachain interactions alone. 
It can find, however, a natural explanation in the framework 
of the extended quasi-1D electron gas model, for which 
interchain interactions can boost the amplitude of CDW 
correlations besides those of the SDW channel that are 
kept essentially unchanged.
 
In the case of non-centrosymmetric anions (e.~g. X=ClO$_4$), our approach
should be refined in order to take into account the doubling of the unit cell
in the transverse direction due to the anion ordering taking place below 24 K
in the normal phase. The concomittant reduction of the Brillouin zone yields 
two electronic bands at the Fermi level and in turn multiple nesting
vectors.\cite{Zanchi01b,Sengupta02} An accurate description of this Fermi
surface should be incorporated in the RG approach, and a modification of the
node structure for the SC gap is expected.\cite{Nickel05b} A previous
(simplified) RPA-like calculation has shown that the nodes of a $d$-wave SC
order 
parameter, which are located at $k_\perp=\pm\pi/2$, are precisely found
where a gap opens due to ClO$_4$ anion ordering, thus making the SC phase
effectively nodeless at low temperature.\cite{Shimahara00}  

The results presented in this work may also be relevant 
for the phase diagram of other series of quasi-1D organic 
superconductors. In this matter, the case of the two-chain 
compounds TTF [M(dmit)$_2$]$_2$ ( M=Ni, Pd) is of 
particular interest. These compounds are characterized by 
an incommensurate CDW state that takes place on the M(dmit)$_2$ 
stacks at low pressure.\cite{Canadell89} At some critical 
pressure, they become superconducting.\cite{Brossard86,Brossard89} 
The variation of the critical temperature under pressure has 
been analyzed in detail in the case of M=Pd, which shows close 
similarity with the one found for the Bechgaard and Fabre 
salts.\cite{Brossard89} The temperature scale for the CDW 
instability at low pressure is relatively large, however, 
and owing to the pronounced  anisotropy of the band parameters 
in these compounds,\cite{Kobayashi87,Canadell89} this 
indicates that the interchain interaction is likely to be 
a key coupling in the stabilization of a 3D ordered 
state in these materials. According to our model, these 
conditions would be favorable to the existence of a 
triplet SC$f$ state in these systems under 
pressure (see \textit{e.g.\ }Fig.~\ref{diagphas}). 

Another result that is highlighted by our analysis in the 
incommensurate case is the occurrence of a singlet SC$g$ state 
when the interchain electron scattering dominates 
for small momentum transfer. The $g_2^\perp$ coupling
tends to suppress  electron pairing between the first 
nearest-neighbor chains, which therefore suppresses 
SC$d$ or SC$f$ type of superconductivity. However, 
an instability of the normal state remains possible. It 
results from the oscillating tail of density-wave 
correlations in the transverse direction which favors longer 
range pairing between electrons separated by more than one 
interchain distance. The fact that an instability of the 
normal state persists in the Cooper channel illustrates 
how the quasi-1D geometry for electrons, with its inherent 
interference between Peierls and Cooper channels, is prone 
to magnify the Kohn-Luttinger mechanism for 
the (Cooper) instability of a Fermi liquid for repulsive 
interactions.\cite{Kohn65} One can easily infer from our 
results that the addition of longer range interchain 
interactions will frustrate short-range interchain pairing and shift it to
larger interchain distances, thus unfolding 
possibilities of superconductivity at even larger 
angular momentum pairing.

\begin{acknowledgments}
J.~C.~N. is grateful to  the Gottlieb Daimler- und Karl Benz-Stiftung for
partial support.  C.~B.  thanks  D. J\'erome, Y. Fuseya, M. Tsuchiizu, 
Y. Suzumura, L.~G. Caron  and 
S. Brown  for  useful discussions and comments, and  the Natural
Sciences and
Engineering Research Council of Canada (NSERC), le Fonds Qu\'ebecois de la
Recherche sur la Nature et les Technologies du Gouvernement du Qu\'ebec
(FQRNT),  and the
Institut Canadien de Recherches Avanc\'ees (CIAR)  for
financial support. N.~D. thanks the condensed matter theory group of the
Universit\'e de Sherbrooke for its hospitality.  
\end{acknowledgments}

\begin{appendix} 

\section{\label{glims}RG equations in limiting cases} 

In order to obtain the one-dimensional limit of the RG equations given
in section~\ref{renorm}, we neglect interchain hopping
($t_\perp=t_\perp'=0$) and take into account intrachain and
nearest-neighbor-chain interactions only. We thus 
obtain\cite{Gorkov74,Lee77} 
\begin{eqnarray} 
\dot{\tilde{\vertex}}_1^{(1D)} & = & 
- \bigl( \tilde{\vertex}_1^{(1D)} \bigr)^2 
- 2 \bigl[ (\tildevertnn_1)^2 + (\tildevertnn_3)^2 \bigr] \ , \\ 
\nonumber 
\dot{\tilde{\vertex}}^{\bot}_1 & = & \tildevertnn_1 \bigl[
-2 \tilde{\vertex}_1^{(1D)} + \tilde{\vertex}_2^{(1D)} 
- \tildevertnn_2 \bigr] 
- \tilde{\vertex}_3^{(1D)} \tildevertnn_3 \ , \\ 
\nonumber 
\dot{\tilde{\vertex}}_2^{(1D)} & = &
- \frac{1}{2} \bigl( \tilde{\vertex}_1^{(1D)} \bigr)^2 
+ \frac{1}{2} \bigl( \tilde{\vertex}_3^{(1D)} \bigr)^2 \ , \\ 
\nonumber 
\dot{\tilde{\vertex}}^{\bot}_2 & = &
- \frac{1}{2} \bigl( \tildevertnn_1 \bigr)^2 
+ \frac{1}{2} \bigl( \tildevertnn_3 \bigr)^2 \ , \\ 
\nonumber 
\dot{\tilde{\vertex}}_3^{(1D)} & = &
\tilde{\vertex}_3^{(1D)} \bigl[ 
- \tilde{\vertex}_1^{(1D)} + 2 \tilde{\vertex}_2^{(1D)} \bigr] 
- 4 \tildevertnn_1 \tildevertnn_3 \ , \\ 
\nonumber 
\dot{\tilde{\vertex}}^{\bot}_3 & = & \tildevertnn_3 \bigl[
-2 \tilde{\vertex}_1^{(1D)} + \tilde{\vertex}_2^{(1D)} 
+ \tildevertnn_2 \bigr] 
- \tilde{\vertex}_3^{(1D)} \tildevertnn_1 \ . 
\end{eqnarray} 

\begin{figure} 
\includegraphics[width=7.0cm]{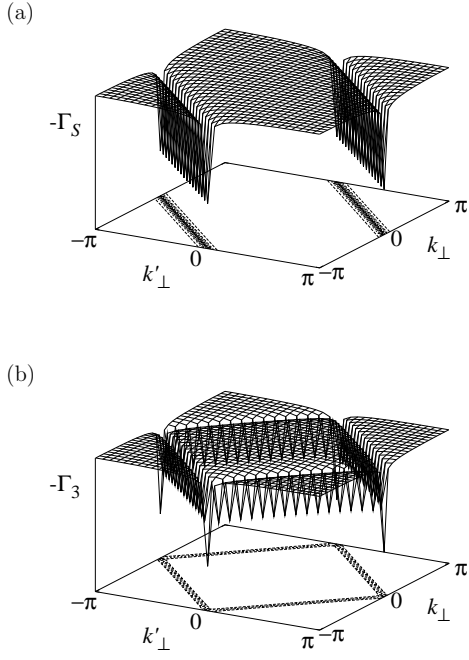} 
\caption{\label{sympeilbild}Typical pictures of the interactions 
$\vertex_S(-k\uperp',k\uperp',k\uperp,-k\uperp)$ (a) and 
$\vertex_3(-k\uperp',k\uperp',k\uperp,-k\uperp)$ (b), 
for initial values $\tilde{g}_S=\tilde{g}_C=0.2$, $\tilde{g}_3=0.02$, 
if the renormalization is restricted to the Peierls channel. 
The result for $\vertex_C$ is identical to that for $\vertex_S$.} 
\end{figure} 

Alternatively, we may restrict the renormalization to a given
correlation channel in certain situations. 
For the particle-hole channel, it is convenient to rewrite the RG
equations in terms of the vertex functions $\vertex_{C,S}$ and 
$\vertex_3^{(C,S)}$, as defined in eqs.~(\ref{CSstvertexdef}). 
Remember however that $\vertex_3^{(C)}$
and $\vertex_3^{(S)}$ are not independent: 
\begin{widetext} 
\[ 
\vertex_3^{(C)}(k_{1\bot}',k_{2\bot}',k_{2\bot},k_{1\bot}) 
= 
- 2 \vertex_3^{(S)}(k_{1\bot}',k_{2\bot}',k_{1\bot},k_{2\bot}) 
+ \vertex_3^{(S)}(k_{1\bot}',k_{2\bot}',k_{2\bot},k_{1\bot}) \ . 
\] 
We obtain 
\begin{eqnarray} 
\dot{\tilde{\vertex}}_C(k_{1\bot}',k_{2\bot}',
k_{2\bot},k_{1\bot})\bigr|_{\Peierls} 
& = & \frac{1}{N\uperp} \sum_{k\uperp} 
B_{\Peierls}(k\uperp,k_{1\bot}'-k_{2\bot}) \\ 
\nonumber && \times \bigl\{ 
\tilde{\vertex}_C(k_{1\bot}',k\uperp,k_{2\bot},k\uperp') 
\tilde{\vertex}_C(k\uperp',k_{2\bot}',k\uperp,k_{1\bot}) \\ 
\nonumber && + \tilde{\vertex}_3^{(C)}(k_{1\bot}',k\uperp,k_{2\bot},k\uperp') 
\tilde{\vertex}_3^{(C)}(k\uperp',k_{2\bot}',k\uperp,k_{1\bot})
\bigr\} \ , 
\end{eqnarray} 
\begin{eqnarray} 
\dot{\tilde{\vertex}}_S(k_{1\bot}',k_{2\bot}',
k_{2\bot},k_{1\bot})\bigr|_{\Peierls} 
& = & \frac{1}{N\uperp} \sum_{k\uperp} 
B_{\Peierls}(k\uperp,k_{1\bot}'-k_{2\bot}) \\ 
\nonumber && \times \bigl\{ 
\tilde{\vertex}_S(k_{1\bot}',k\uperp,k_{2\bot},k\uperp') 
\tilde{\vertex}_S(k\uperp',k_{2\bot}',k\uperp,k_{1\bot}) \\ 
\nonumber && + \tilde{\vertex}_3^{(S)}(k_{1\bot}',k\uperp,k_{2\bot},k\uperp') 
\tilde{\vertex}_3^{(S)}(k\uperp',k_{2\bot}',k\uperp,k_{1\bot})
\bigr\} \ , 
\end{eqnarray} 
\begin{eqnarray} 
\dot{\tilde{\vertex}}_3^{(S)}(k_{1\bot}',k_{2\bot}',k_{2\bot},k_{1\bot})
\bigr|_{\Peierls} 
& = & \frac{1}{N\uperp} \sum_{k\uperp} 
B_{\Peierls}(k\uperp,k_{1\bot}'-k_{1\bot}) \\ 
\nonumber && 
\times \frac{1}{2} \times \bigl\{ 
- \tilde{\vertex}_C(k_{1\bot}',k\uperp,k_{1\bot},k\uperp') 
\tilde{\vertex}_3^{(C)}(k\uperp',k_{2\bot}',k\uperp,k_{2\bot}) \\ 
\nonumber && 
- \tilde{\vertex}_3^{(C)}(k_{1\bot}',k\uperp,k_{1\bot},k\uperp') 
\tilde{\vertex}_C(k\uperp',k_{2\bot}',k\uperp,k_{2\bot}) \\ 
\nonumber && 
+ \tilde{\vertex}_S(k_{1\bot}',k\uperp,k_{1\bot},k\uperp') 
\tilde{\vertex}_3^{(S)}(k\uperp',k_{2\bot}',k\uperp,k_{2\bot}) \\ 
\nonumber && 
+ \tilde{\vertex}_3^{(S)}(k_{1\bot}',k\uperp,k_{1\bot},k\uperp') 
\tilde{\vertex}_S(k\uperp',k_{2\bot}',k\uperp,k_{2\bot}) \bigr\} \\ 
\nonumber && 
+ \frac{1}{N\uperp} \sum_{k\uperp} 
B_{\Peierls}(k\uperp,k_{1\bot}'-k_{2\bot}) \\ 
\nonumber && \times \bigl\{ 
\tilde{\vertex}_S(k_{1\bot}',k\uperp,k_{2\bot},k\uperp') 
\tilde{\vertex}_3^{(S)}(k\uperp',k_{2\bot}',k\uperp,k_{1\bot}) \\ 
\nonumber && 
+ \tilde{\vertex}_3^{(S)}(k_{1\bot}',k\uperp,k_{2\bot},k\uperp') 
\tilde{\vertex}_S(k\uperp',k_{2\bot}',k\uperp,k_{1\bot})
\bigr\} \ . 
\end{eqnarray} 
These equations are remarquably symmetric with respect to spin and
charge density correlations, cf.\ \textit{e.g.\ }the results when we
take $g_S=g_C$ from the beginning (Fig.~\ref{sympeilbild}). 
For low temperature and energy cutoff $\Lambda$, the particle-hole
loop integral $B_{\Peierls}(q\uperp)$ is strongly peaked for momentum
transfers equal to the best nesting vector, 
\textit{i.~e.\ }$q\uperp=\pi$. 
As can be seen from the preceding equations, this
generates a peak at $k_{1\bot}'-k_{2\bot}=\pi$ in $\vertex_S$
(responsible for the SDW) and in $\vertex_C$
(responsible for the CDW). In $\vertex_3$, peaks are
created at $k_{1\bot}'-k_{2\bot}=\pi$ as well as 
$k_{1\bot}'-k_{1\bot}=\pi$. The peak in $\vertex_3$ for
$k_{1\bot}'-k_{2\bot}=\pi$ enhances the spin 
density correlations, whereas the one for $k_{1\bot}'-k_{1\bot}=\pi$
supports the charge density correlations. 

Let us now consider the influence of interchain interactions 
[Eq.~(\ref{gstarting})] on these structures. 
Interchain Umklapp scattering $\gnn_3$ 
reduces the latter peak and thus weakens the
charge fluctuations, whereas it has no direct influence on the spin
correlations. $\gnn_1$ and $\gnn_2$ have no direct effect on the spin
fluctuations either. Interchain backward scattering $\gnn_1$, on the
contrary, reinforces the peak in $g_C$ at $k_{1\bot}'-k_{2\bot}=\pi$ and
thus supports the formation of charge density waves. 

We finally give the RG equations restricted to the particle-particle
channel. They are most conveniently written in terms of 
the singlet and triplet pair interactions, $\alpha=s$, $t$: 
\begin{eqnarray} 
\dot{\tilde{\vertex}}_{\alpha}(k_{\bot 1}'k_{\bot 2}'k_{\bot 2}k_{\bot 1}) 
\bigr|_{\Cooper} & = & 
\frac{1}{N\uperp} \sum_{k\uperp} 
B_{\Cooper}(k\uperp,\qcooperperp) \\ 
\nonumber && \times 
\tilde{\vertex}_{\alpha}(k_{\bot 1}'k_{\bot 2}'k\uperp k\uperp') 
\tilde{\vertex}_{\alpha}(k\uperp k\uperp'k_{\bot 1}k_{\bot 2}) 
\biggr|_{\zweikleinemathezeilen{
k\uperp'=-k\uperp+\qcooperperp}{
\qcooperperp=k_{\bot 1}+k_{\bot 2}} } \ . 
\end{eqnarray} 
\end{widetext}

\end{appendix}


\begin{thebibliography}{81}
\bibitem[{not()}]{note1}
\bibinfo{note}{In this paper, we do not consider the case $g_1\simeq g_2\simeq
  g_3$ (Hubbard model) which is characterized by a strong-coupling fixed point
  in the charge sector describing a 1D Mott insulator with long-range
  antiferromagnetic order at low temperatures due to interchain exchange
  interaction.\cite{Bourbon95}}.
\end{thebibliography}

\end{document}